\documentclass[usenatbib]{mn2e}
\usepackage{multirow}
\usepackage{colortbl}
\usepackage{color}
\usepackage[fleqn]{amsmath}
\usepackage{amssymb}
\usepackage{amsfonts} 
\usepackage{verbatim}
\usepackage{scalefnt}
\usepackage{lscape}
\usepackage{booktabs}
\usepackage{float}
\usepackage{multirow}
\usepackage{adjustbox}
\usepackage{rotating}
\usepackage[dvipsnames]{xcolor}
\usepackage{comment}
\usepackage{siunitx}
\usepackage{scalerel,amssymb}
\usepackage{ulem}

\usepackage[bookmarks=False,bookmarksnumbered,colorlinks=true, citecolor=blue, linkcolor=cyan]{hyperref}

\newcommand{\apjl}{ApJ}
\newcommand{\apjs}{ApJS}
\newcommand{\ao}{Appl.~Opt.}
\newcommand{\aap}{A\&A}

\newcommand{\beq}{\begin{equation}}
\newcommand{\eeq}{\end{equation}}
\usepackage{longtable}

\title[BHs and AGN in low-mass galaxies]{The black hole population in low-mass galaxies in large-scale cosmological simulations}
\author[Haidar \& Habouzit]{Houda Haidar$^{1}$\thanks{E-mail: houda.haidar@obspm.fr}, M\'{e}lanie Habouzit$^{2,3}$, Marta Volonteri$^1$, Mar Mezcua$^{4,5}$, 
\newauthor
Jenny Greene$^6$, Nadine Neumayer$^3$, Daniel Angl\'es-Alc\'azar$^{7,8}$, 
\newauthor
Ignacio Martin-Navarro$^{9,10}$, Nils Hoyer$^3$, Yohan Dubois$^1$, Romeel Dav\'{e}$^{11}$
\\
 $^1$ Institut d'Astrophysique de Paris, IAP, Sorbonne Universit\'es, CNRS, UMR 7095, 98 bis bd Arago, 75014 Paris, France\\
 $^2$ Zentrum für Astronomie der Universit\"at Heidelberg,
 ITA, Albert-Ueberle-Str. 2, D-69120 Heidelberg, Germany\\
 $^3$ Max-Planck-Institut f\"ur Astronomie, MPIA, K\"onigstuhl 17, D-69117 Heidelberg, Germany\\
 $^4$ Institute of Space Sciences (ICE, CSIC), Campus  UAB, Carrer de Magrans, 08193 Barcelona, Spain \\
$^5$ Institut d’Estudis Espacials de Catalunya (IEEC), Carrer Gran Capità, 08034 Barcelona, Spain \\
 $^6$ Department of Astrophysical Sciences, Princeton University, 4 Ivy Lane, Princeton, NJ, 08544, USA\\
 $^{7}$ Department of Physics, University of Connecticut, 196 Auditorium Road, U-3046, Storrs, CT 06269-3046, USA\\
 $^8$ Center for Computational Astrophysics, Flatiron Institute, New York, NY 10010, USA\\
 $^9$ Instituto de Astrof\'isica de Canarias, V\'ia L\'actea s/n, E-38205 La Laguna, Tenerife, Spain \\
$^{10}$ Departamento de Astrof\'isica, Universidad de La Laguna, E-38205 La Laguna, Tenerife, Spain\\
 $^{11}$ Institute for Astronomy, Royal Observatory, University of Edinburgh, Edinburgh EH9 3HJ, UK\\
}

\date{Accepted 2022 June 6. Received 2022 May 24; in original form 2022 January 24}                 

\begin{document}
\maketitle

\begin{abstract}
Recent systematic searches for massive black holes (BHs) in local dwarf galaxies led to the discovery of a population of faint Active Galactic Nuclei (AGN). 
We investigate the agreement of the BH and AGN populations in the Illustris, TNG, Horizon-AGN, EAGLE, and SIMBA simulations with current observational constraints in low-mass galaxies. 
We find that some of these simulations produce BHs that are too massive, and that the BH occupation fraction at $z=0$ is not inherited from the simulation seeding modeling.
The ability of BHs and their host galaxies to power an AGN depends on BH and galaxy subgrid modeling. The fraction of AGN in low-mass galaxies is not used to calibrate the simulations, and thus can be used to differentiate galaxy formation models. AGN fractions at $z=0$ span two orders of magnitude at fixed galaxy stellar mass in simulations, similarly to observational constraints, but uncertainties and degeneracies affect both observations and simulations. 
The agreement is difficult to interpret due to differences in the masses of simulated and observed BHs, BH occupation fraction affected by numerical choices, and an unknown fraction of obscured AGN.
Our work advocates for more thorough comparisons with observations 
to improve the modeling of cosmological simulations, and our understanding of BH and galaxy physics in the low-mass regime. The mass of BHs, their ability to efficiently accrete gas, and the AGN fraction in low-mass galaxies have important implications for the build-up of the entire BH and galaxy populations with time.
\end{abstract}

\section{Introduction}

Over the last decade, compelling evidence has shown that massive black holes (BHs) are commonly found in the center of galaxies in the local Universe, and that their masses correlate with e.g. the stellar mass of their host galaxies or their bulges \citep[e.g.,][]{2009ApJ...698..198G}. 
Today, there is a systematic search for the presence of BHs in low-mass galaxies of $M_{\star}\leqslant 10^{9.5}\, \rm M_{\odot}$, with a focus on accreting BHs -commonly referred to as Active Galactic Nuclei (AGN)-, using multi-wavelength observations
\citep{Greene2008,reines2015relations,2016PASA...33...54R,2016ApJ...817...20M,2017IJMPD..2630021M,2018ApJ...863....1C,2019arXiv191109678G,2020ApJ...888...36R}. The least massive BHs observed 
are the BH found in RGG118 with a mass of $5\times 10^{4}\, \rm M_{\odot}$   \citep{2015ApJ...809L..14B}, and 
the BH located in NGC4395 \citep{1989ApJ...342L..11F} with a BH mass estimate in the range of $\sim  9100{-}4\times 10^{5}\, \rm M_{\odot}$ \citep{2019NatAs...3..755W}. Several techniques are used to detect BHs and AGN  in dwarf galaxies, and include optical spectroscopy, observations in X-ray, radio, and even infrared (IR) 
\citep[e.g.,][]{2004ApJ...610..722G,Greene2008,2012ApJ...755..167D,2018ApJ...863....1C,2018ApJS..235...40L,2018MNRAS.478.2576M,2019MNRAS.488..685M,2019MNRAS.489L..12K}. Multi-wavelength observations are often used to confirm candidates, as for example optical tracers in young starburst regions in low-metallicity environment can resemble those of AGN. Winds from both supernovae (SN) and massive stars are also a source of contamination.
Many AGN are detected based on their X-ray emission, but the presence of X-ray binaries (XRBs) can make the interpretation difficult for faint AGN \citep[e.g.,][]{2008ApJ...680..154G,2010ApJ...714...25G,reines2011actively,2016ApJ...817...20M,2016ApJ...831..203P,2017ApJ...842..131S,2018MNRAS.478.2576M,2018ApJ...865...43F,2020arXiv201102501S}. 
In the low-mass regime, the ubiquity of BHs is not unique to the center of galaxies. \citet{2020ApJ...888...36R} discovered that half a dozen of dwarf galaxies were hosting offset radio sources that could be AGN. 
Faint off-center AGN signatures\footnote{ Observational signatures of off-center AGN are in some cases also consistent with light echoes of past AGN activity \citep{mezcua2020hidden}.} have been also recently detected in dwarf galaxies with MaNGA integral field unit (IFU)
observations \citep{2020ApJ...898L..30M}, confirming that the search for BHs should not be limited to central AGN.

The mechanisms of BH formation are still unknown, but the BH population in local dwarf galaxies provides us with indirect constraints \citep{2008MNRAS.383.1079V,2010MNRAS.408.1139V}. 
One of the leading mechanisms forms BHs as remnants of the first generation of stars with initial masses of $M_{\rm BH}\sim 100\, \rm M_{\odot}$ \citep[``PopIII remnant model''][]{MadauRees2001,volonteri2003a,JBromm,Johnson2007}. BHs could also form in slightly metal-enriched stellar compact clusters where runaway stellar collisions can trigger the formation of a very massive star that collapse onto a BH of $M_{\rm BH}\sim 10^{2} - 10^{4}\, \rm M_{\odot}$ \citep[``Compact stellar cluster model''][]{2008ApJ...686..801O,
Regan09,2020MNRAS.493.2352A}. 
In the ``direct collapse'' model \citep[e.g.,][]{Begelman2006,BVR2006}, a large amount of gas collapses isothermally without fragmenting and forms a supermassive star collapsing into a 
BH of $M_{\rm BH}\sim 10^{4}-10^{6}\,\rm M_{\odot}$. This model requires several key conditions such as the absence of molecular hydrogen coolant to avoid gas fragmentation. 
This can be achieved by the presence of a high Lyman-Werner radiation (11.2-13.6 eV photons) coming from nearby star-forming regions \citep{2008MNRAS.391.1961D,2014MNRAS.442L.100V,2014MNRAS.443..648A,2016arXiv160100557H,2021MNRAS.502..700C},
but is not necessary in the case of dynamical heating in the host halo history \citep{2019Natur.566...85W}.

The PopIII remnant and compact cluster models predict the formation of BHs in great abundance, inhabiting almost  -if not all-  galaxies in the high-redshift Universe \citep[e.g.,][]{2012NatCo...3E1304G,devecchi2009formation}.
However, if BHs form through less common processes such as in the direct collapse model, only few galaxies could host BHs. 
While it is currently impossible to constrain the fraction of galaxies hosting a BH at redshifts close to those of BH formation,  we can obtain indirect constraints in local low-mass galaxies using {\it archaeology}. These constraints include the mass of BHs in dwarf galaxies and the BH occupation fraction in these galaxies.
Theoretically, the assembly and merger history of low-mass galaxies through cosmic time is predicted to be quieter than for more massive galaxies. First, the mass of BHs in dwarf galaxies could  be close to their seeding mass. Second, dwarf galaxies with a possible low number of mergers in their history could provide us with constraints on the intrinsic occupation fraction of the high-redshift Universe.

Constraining the galaxy BH occupation fraction in observations is challenging. The few existing constraints often rely on extrapolations from the AGN occupation fraction. BHs in dwarf galaxies have low masses, and thus are more likely to power faint AGN, which are difficult to identify because of the numerous sources of contamination described above.
Current observational constraints indicate that more than $50\%$ of galaxies with $M_{\star}\leqslant 10^{10}\, \rm M_{\odot}$ could host a BH \citep{2015ApJ...799...98M,trump2015biases,2019ApJ...872..104N}, with large uncertainties and differences among these constraints. Assessing the galaxy BH occupation fraction theoretically is also difficult. This is because capturing the physics of BH formation and growth and galaxy evolution
in a cosmological context is very demanding and requires high resolution simulations to capture low-mass galaxies 
and follow their evolution from the redshift of BH formation to the local Universe.

The BH occupation fraction for different BH formation mechanisms was first investigated with semi-analytical models \citep{VHM,VLN2008}. Later, \citet{Bellovary2011} used zoom-in cosmological simulations of low-mass halos to derive the number of low-mass galaxies that could host a BH in the local Universe. The occupation fraction was studied 
with large-scale cosmological simulations in \citet{2017MNRAS.468.3935H}, from high-redshift down to $z=2$. These simulations include a physical BH formation model mimicking the PopIII remnant and compact stellar cluster models. 
The AGN occupation fraction was investigated in the {\sc FABLE} cosmological simulation (whose BH seeding model does not follow one of the theoretical models) down to $z=0$ in \citet{2021MNRAS.503.3568K}. Recently, a model for the AGN occupation fraction was derived in 
\citet{2021ApJ...920..134P}, based on gas properties and angular momentum of galaxies and calibrated against the X-ray AGN occupation fractions. 

Cosmological simulations are a great tool to study the evolution of massive BHs, as well as their interplay with their host galaxies. 
The rise of large-scale simulations such as Illustris \citep{2014MNRAS.445..175G,2014MNRAS.444.1518V,2015MNRAS.452..575S,2015A&C....13...12N}, TNG100, TNG300 \citep{2017arXiv170302970P,2017arXiv170703397S,2018MNRAS.475..624N,2017arXiv170703401N,2018MNRAS.480.5113M}, Horizon-AGN \citep{2014MNRAS.444.1453D,2016MNRAS.463.3948D,2016MNRAS.460.2979V}, EAGLE  \citep{2015MNRAS.446..521S,2015MNRAS.450.1937C,2016A&C....15...72M} and SIMBA \citep{2019MNRAS.486.2827D,2019MNRAS.487.5764T}
allows us to numerically bridge the gap between observational and theoretical studies. Such simulations have been successful at producing galaxies in good agreement with the properties of observed galaxies.
Nevertheless, discrepancies among the simulations exist and arise
from the use of different galaxy and BH subgrid physics 
\citep[]{2021MNRAS.503.1940H}.
The formation of BHs in these simulations does not follow the theoretical prescriptions of BH formation mechanisms. Nevertheless, as we will see in this paper, BH activity in low-mass galaxies can provide us with key hints on the interplay between BHs and their host galaxies in this regime.  In particular, it can constrain the modeling of seeding, accretion, and feedback processes from SNe and AGN in simulations.

In this work, we analyze the presence of BHs and AGN, and their properties, in the dwarf galaxies of six state-of-the-art cosmological simulations.
Throughout this paper, we refer to the low-mass regime as galaxies with $ 10^{9} \leqslant M_{\star}/\rm M_{\odot}  \leqslant 10^{10.5} $ and dwarf galaxies as $ 10^{9}  \leqslant M_{\star}/\rm M_{\odot} \leqslant 10^{9.5}$. 
We describe the simulations in Section \ref{sec:2}. We compute the BH occupation fractions produced by the six simulations in Section \ref{sec:3}. In Section \ref{sec:4}, we derive the AGN occupation fractions of the simulations and assess their agreement with current observational constraints. In Section \ref{sec:5}, we examine the properties of the AGN host galaxies. 
Discussion and conclusion points are given in Section \ref{sec:6} and \ref{sec:7}. 

\section{Methodology: Cosmological simulations}
\label{sec:2}
Cosmological simulations follow the evolution of the dark matter and baryonic matter contents in an expanding space-time, and allow us to study the evolution of thousands of galaxies with a broad range of total stellar masses from $M_{\star}=10^{9}$ to $10^{12}\, \rm M_{\odot}$. In this paper, we analyze the six Illustris, TNG100, TNG300 (larger volume and lower resolution with respect to TNG100), Horizon-AGN, EAGLE, and SIMBA simulations.
They all have volumes of $\geqslant 100^{3}\, \rm cMpc^{3}$, dark matter mass resolutions of $\sim 5\times 10^{6}-8\times 10^{7}\, \rm M_{\odot}$, baryonic resolutions of $10^{6} - 2 \times 10^{7}\, \rm M_{\odot}$, and spatial resolutions of $\sim 1-2$ ckpc. 
Physical processes taking place at small scales below the galactic scale are modeled as subgrid physics. This includes e.g., star formation, stellar and SN feedback, BH formation, evolution and feedback. Subgrid models vary from simulation to simulation \citep[see][for a summary of all these simulations]{2021MNRAS.503.1940H}. Detailed descriptions of the simulation's BH modelings can be found in \citet{2014MNRAS.445..175G,2014MNRAS.444.1518V} for Illustris, \citet{2017arXiv170302970P,2018MNRAS.479.4056W} for TNG, \citet{2016MNRAS.463.3948D,2016MNRAS.460.2979V} for Horizon-AGN, \citet{2015MNRAS.446..521S,2016MNRAS.462..190R,2018MNRAS.481.3118M,2017MNRAS.468.3395M} for EAGLE, and \citet{2019MNRAS.486.2827D,2019MNRAS.487.5764T,2021MNRAS.503.3492T} for SIMBA.

\subsection{Modeling of BH physics}
The modeling of BH seeding, accretion and feedback are defined differently across simulations.
Once the BH seed is planted, it can start growing through feeding on its surrounding gas and BH mergers. 
In cosmological simulations, BH accretion is often modeled with the Bondi model or a variation of it. For example, the accretion rate onto TNG BHs includes a magnetic field component as the TNG simulations are magneto-hydrodynamical simulations \citep{2017arXiv170302970P}. Moreover, EAGLE includes a viscous disk component \citep{2015MNRAS.454.1038R}, while SIMBA incorporates two modes of accretion: the Bondi model for the hot gas component ($T>10^{5}\, \rm K$) and a gravitational torque limited model for the cold gas \citep[$T<10^{5}\, \rm K$,][]{2011MNRAS.415.1027H,2015ApJ...800..127A,2017MNRAS.464.2840A}.
All simulations rely on AGN feedback to produce a population of quenched massive galaxies, but each simulation has its own implementation  
and strength of the feedback. Simulations 
either use a single-mode feedback as EAGLE \citep{2015MNRAS.446..521S}, or a two-mode feedback for the others \citep{2015MNRAS.452..575S,2017MNRAS.465.3291W,2016MNRAS.463.3948D,2019MNRAS.486.2827D}. The released energy can be e.g., thermal and isotropic, and/or kinetic with collimated jets, or non-collimated outflows. 

In this work, we are particularly interested in BH seed modeling. 
BHs are seeded in massive halos of $7 \times 10^{10}\, \rm M_{\odot}$ in Illustris, TNG100, and TNG300, and in galaxies of $M_{\star}\geqslant 10^{9.5}\, \rm M_{\odot}$ in SIMBA. 
Notwithstanding this condition, SIMBA shows a small fraction of galaxies with $M_{\star}\leqslant 10^{9.5}\, \rm M_{\odot}$ harboring a BH. These can be seeded galaxies that lost mass due to interactions and stripping processes. 
These galaxies can also be the result of approximate mass measurement from the on-the-fly friends-of-friends algorithm used to identify galaxies for the seeding; the mass of these galaxies would be larger than the post-processing masses. We do not consider these galaxies in our analysis. In Horizon-AGN, the location of BH formation is based on the properties of the local gas cells instead of the halo or galaxy mass.
Initial BH masses are fixed and comprised in the range $M_{\rm BH}=10^{4}-10^{6}\, \rm M_{\odot}$, with $\sim 10^{4}\, \rm M_{\odot}$ for SIMBA, $\sim 10^{5}\, \rm M_{\odot}$ for Illustris, Horizon-AGN, EAGLE, and $\sim 10^{6}\, \rm M_{\odot}$ for TNG100, and TNG300. These simulations do not employ one of the theoretical BH formation models, such as the PopIII remnant or direct collapse models. The initial mass of BHs and the prescriptions for seeding are instead often chosen to reproduce one of the BH to stellar mass
empirical scaling relations observed in the local Universe.

For what follows, we only consider galaxies with stellar mass $M_{\star}\geqslant 10^{9}\, \rm M_{\odot}$. It is also important to mention that we do not distinguish between central and satellite galaxies. Among all the simulations, Horizon-AGN is the only one that forms BHs based on the local properties of the gas. There is an exclusion radius of 50 ckpc preventing the formation of a BH in a galaxy that already hosts one. In Horizon-AGN, galaxies can have multiple BHs as a result of galaxy mergers. 
When this occurs, we only consider the most massive BH within the galaxies. Two BHs merge if their separation is smaller than the spatial resolution of the simulation \citep[][]{2014MNRAS.444.1453D}. 
In most simulations, BHs are re-positioned to the galaxy potential well at every time step, which means that these simulations can not be used to assess the number of wandering BHs in galaxies.

\subsection{AGN luminosity}
We follow the model of \citet{Churazov2005} to compute the AGN luminosity. As such, we explicitly differentiate between radiatively efficient  
and inefficient AGN. 
For radiatively efficient AGN with Eddington ratios of $f_{\rm Edd}=\dot{M}_{\rm BH}/\dot{M}_{\rm Edd}>0.1$, the bolometric luminosity can be expressed as
$L_{\rm bol}=0.1 
\dot{M}_{\rm BH} c^{2}$,
where $\epsilon_{\rm r}=0.1$ is the radiative efficiency, $\dot{M}_{\rm BH}$ is the accretion rate onto the BH, and $c$ is the speed of light.
Radiatively inefficient AGN with $f_{\rm Edd}\leqslant 0.1$ have a bolometric luminosity defined as:
\begin{eqnarray}
L_{\rm bol}=0.1 L_{\rm Edd} (10 f_{\rm Edd})^{2}=(10 f_{\rm Edd}) \epsilon_{\rm r} \dot{M}_{\rm BH} c^{2}.
\end{eqnarray}
 We caution that we lowered the radiative efficiency assumed in the BH accretion and AGN feedback models of the Illustris and TNG simulations. 
Taking the same radiative efficiency of 0.1 in the post-processing analysis of the simulations is motivated by a better agreement with observational constraints on the faint end of the AGN luminosity function for Illustris and TNGs \citep{2022MNRAS.509.3015H}.
This choice
does not affect the conclusions of the present paper. 
 For consistency with previous simulation papers, we compute AGN hard X-ray luminosities by applying the bolometric correction of \citet{Hop_bol_2007}. As tested in \citet{2022MNRAS.509.3015H} the correction derived in 
\citet{2020A&A...636A..73D}, which is very slightly lower than the one derived in \citet{Hop_bol_2007} in the low-mass galaxy regime, does not impact our results. For the purpose of this paper we also tested the correction of \citet{2020MNRAS.495.3252S}, which returned similar results as that of \citet{Hop_bol_2007} for $M_{\star} \leqslant 10^{10.5}\, \rm M_{\odot}$.

\subsection{Post-processing modeling of the X-ray binary population}

We follow \citet{2020arXiv201102501S} to parameterize the X-ray emission ($2-10\, \rm keV$ band) of the XRB 
population of the simulated galaxies.
XRBs consist essentially of a normal star and a compact object (either a neutron star or a BH) orbiting each other. Through this process, the material from the star is drawn towards the collapsed object giving rise to 
X-ray emission. The binary system makes up two categories depending on the mass of the star: low-mass XRBs (LMXBs) and high-mass XRBs (HMXBs). The former has a lifetime longer than $1 \, \rm Gyr$ 
and their number in a galaxy is a function of the galaxy stellar mass, 
whereas the latter 
has a shorter lifetime of $< 100 \, \rm Myr$ and as such, will be more numerous in star-forming galaxies.

The galaxy-wide XRB luminosity includes the contributions of both LMXBs and HMXBs, as:
\begin{eqnarray}
\frac{L_{\rm XRB}}{\rm erg/s}=\alpha_{\rm LMXB} \left(1+z \right)^{\gamma} \frac{M_{\star}}{\rm M_{\odot}}+\beta_{\rm HMXB} \left(1+z \right)^{\delta} \rm \frac{SFR}{\rm M_{\odot}/yr},
\label{eq:xrb_relation}
\end{eqnarray}
with $M_{\star}$ the galaxy total stellar mass, SFR the galaxy star formation rate, and  $z$ the galaxy redshift. We employ the parameters of \citet{2019ApJS..243....3L}: $\log_{10}\alpha_{\rm LMXB}=29.15\, \rm erg/s/M_{\odot}$, $\log_{10}\beta_{\rm HMXB}=39.73\, \rm erg/s/(M_{\odot}/yr)$, $\gamma=2.0$, and $\delta=1.3$.
We note here that the empirical relations linking the X-ray luminosity of the galaxy XRB population to the galaxy stellar mass and SFR are derived from relatively massive and nearby galaxies. Here we choose to employ the most recent of such relations, but others have been derived from observations \citep[e.g.,][]{2010ApJ...724..559L,2016ApJ...825....7L,2019ApJS..243....3L,2017MNRAS.465.3390A,2018ApJ...865...43F}. Moreover, the galaxy-wide XRB emission could also depend on the metallicity and stellar ages of the galaxy \citep[e.g.,][]{2013ApJ...776L..31F,2016ApJ...825....7L,2017ApJ...840...39M,2021ApJ...907...17L}, but we neglect these possible dependencies.
In this paper, we extend this empirical relation to galaxies 
with stellar mass $M_{\star}\leqslant 10^{10.5}\, \rm M_{\odot}$. 

\subsection{Post-processing modeling of the X-ray emission from the galaxy ISM}

In the quest for X-ray emission, observers can receive 
from the sky three types of sources: hot gas from the interstellar medium (ISM), AGN and XRBs. The contributions from the latter two are predicted to be the most dominant in the hard X-ray band. 
The hot gas is predicted to be significant in the rest-frame energy $< 1.5\, \rm keV$, whereas XRBs dominate the emission over the hot gas for higher energies of $> 1.5\, \rm keV$ \citep[e.g.,][]{2016ApJ...825....7L}.
We compute the contribution of the ISM to the galaxy-wide total X-ray luminosity for all the simulations and show that the ISM contribution is indeed minor (see Section~\ref{sec:XRAY-contribution}
).

We model the X-ray luminosity of the ISM by adopting the empirical relation of \citet{2012MNRAS.419.2095M} (in the soft 0.5-2 keV band): 

\begin{eqnarray}
\frac{L_{\rm ISM}}{{\rm erg/s}}=8.3\times 10^{38}\times \rm \frac{SFR}{M_{\odot}/yr}.
\end{eqnarray}

\noindent We then employ a k-correction to move from soft band to hard X-ray band , assuming a photon index of $\Gamma=3$ \citep{2018MNRAS.478.2576M}. In this paper, we also use this modeling to estimate the AGN contribution from the galaxy-wide X-ray luminosity provided in some observational studies (see Appendix ~\ref{sec:observations_list}).

\section{Results: BH occupation fraction}
\label{sec:3}

\subsection{Local Universe}
\begin{figure*}
    \centering
    \includegraphics[scale=0.58]{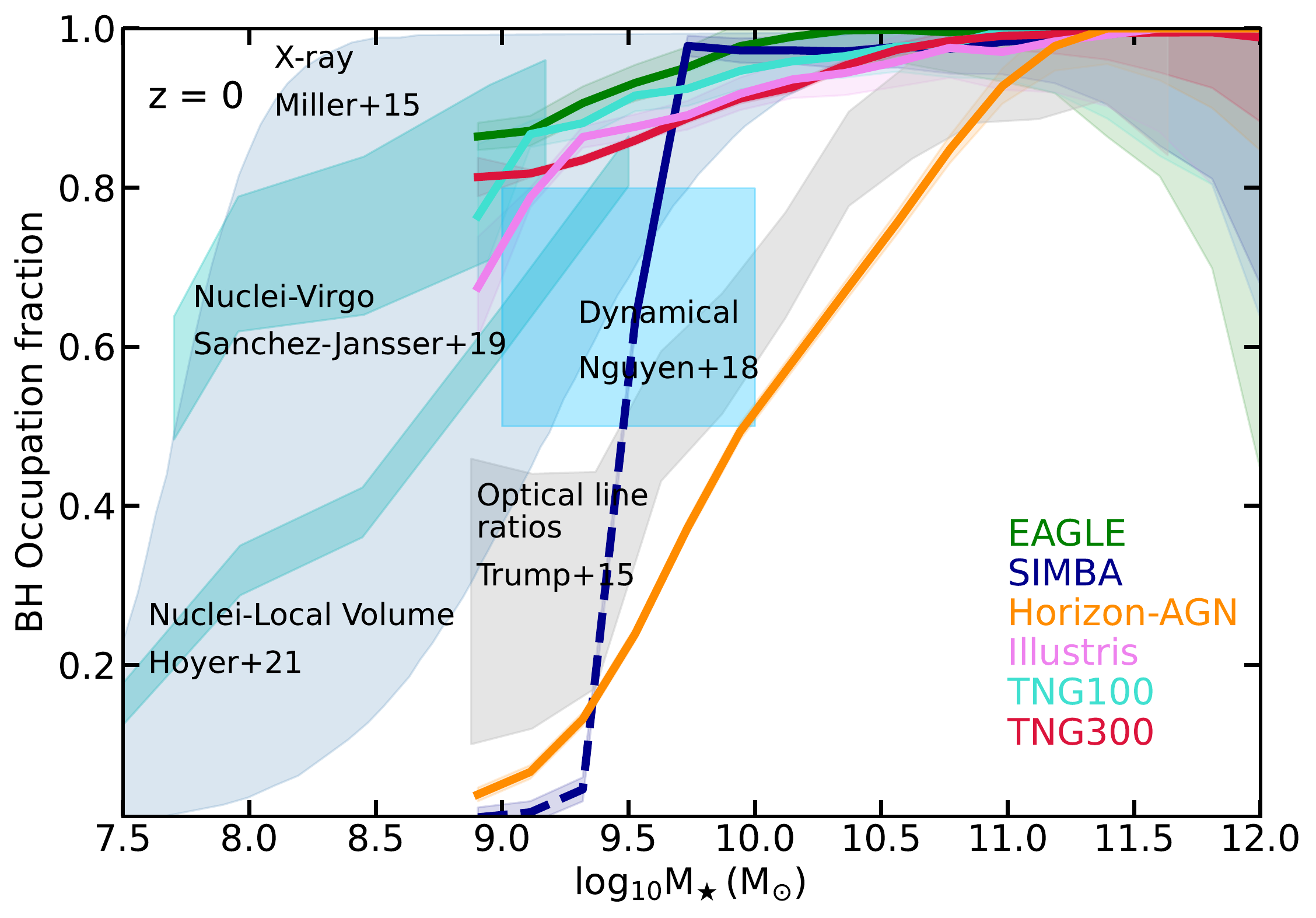}
    \caption{
    The BH occupation fraction for the local Universe ($z=0$) predicted by the six simulations. Poisson error bars are shown as shaded regions, and are larger in more massive galaxies due to the presence of fewer galaxies. We show the fraction of SIMBA as a dashed line for $M_{\star}< 10^{9.5}\, \rm M_{\odot}$, as only more massive galaxies are seeded. 
    We include the current observational constraints on the BH occupation fraction of \citet{2018ApJ...858..118N}
    with dynamical mass measurement of BHs in a few nearby nucleated galaxies, and from \citet{2015ApJ...799...98M} and \citet{trump2015biases}, both derived from a sample of AGN in low-mass galaxies. While not being a constraint on the BH occupation fraction, we also report the galaxy nucleation fraction in the galaxy cluster Virgo \citep{2019ApJ...878...18S} and in the Local Volume \citep{hoyer2021nucleation}.
}
    \label{fig:BHoccfrac-err}
\end{figure*}

The probability of a given galaxy to host a BH is defined by the BH occupation fraction (OF), which is simply derived as the ratio between galaxies hosting a central massive BH and the total number of galaxies, in bins of galaxy stellar mass. We show the BH OF of all the simulations at $z=0$ in Fig.~\ref{fig:BHoccfrac-err}.
The BH OF increases with the galaxy stellar mass for all the simulations. 
While all of them have an occupation of unity in massive galaxies with $M_{\star}\geqslant 10^{11}\, \rm M_{\odot}$, they do not reach unity at the same galaxy stellar mass. The BH OF reaches unity at $M_{\star}=10^{10.5} \rm M_{\odot}$ in EAGLE, TNG100 and TNG300, at $M_{\star}=10^{11.3} \rm M_{\odot}$ in SIMBA and Horizon-AGN and finally, at $M_{\star}=10^{11} \rm M_{\odot}$ in Illustris.

All the simulations except Horizon-AGN have a BH occupation higher than $90\%$ for galaxies with $M_{\star}= 10^{10}\, \rm M_{\odot}$. The difference among simulations extends and increases down to the dwarf galaxy regime. 
The probability of a simulated galaxy with $M_{\star}=10^{9}\, \rm M_{\odot}$ to host a BH is the lowest in Horizon-AGN with $7\%$, and ranges from $71\%$ to $86\%$ for the other simulations. 
The seeding of BHs in galaxies with $ \rm M_{\star}\geqslant 10^{9.5}\, \rm M_{\odot}$ in SIMBA does not allow us to investigate the BH occupation in the dwarf regime with $ \rm M_{\star}\leqslant 10^{9.5}\, \rm M_{\odot}$(shown as dashed line in Fig.~\ref{fig:BHoccfrac-err}). Galaxies of $ \rm M_{\star}\geqslant 10^{9.5}\, \rm M_{\odot}$ in SIMBA are, however, reaching an OF of $90\%$, similar to most of the other simulations. Horizon-AGN, which is the only simulation avoiding a fixed halo/galaxy mass threshold for seeding BHs, has a different behavior: its BH OF is low ($10\%$) in the dwarf regime, and increases progressively to reach $100\%$ only in galaxies with $M_{\star}\geqslant 10^{11.3}\, \rm M_{\odot}$. To understand this feature, we investigate the redshift evolution of the BH OF in Section \ref{sec:sec3.3}. 

\begin{figure*}
    \centering
    \hspace*{-0.9cm}
    \includegraphics[scale=0.3]{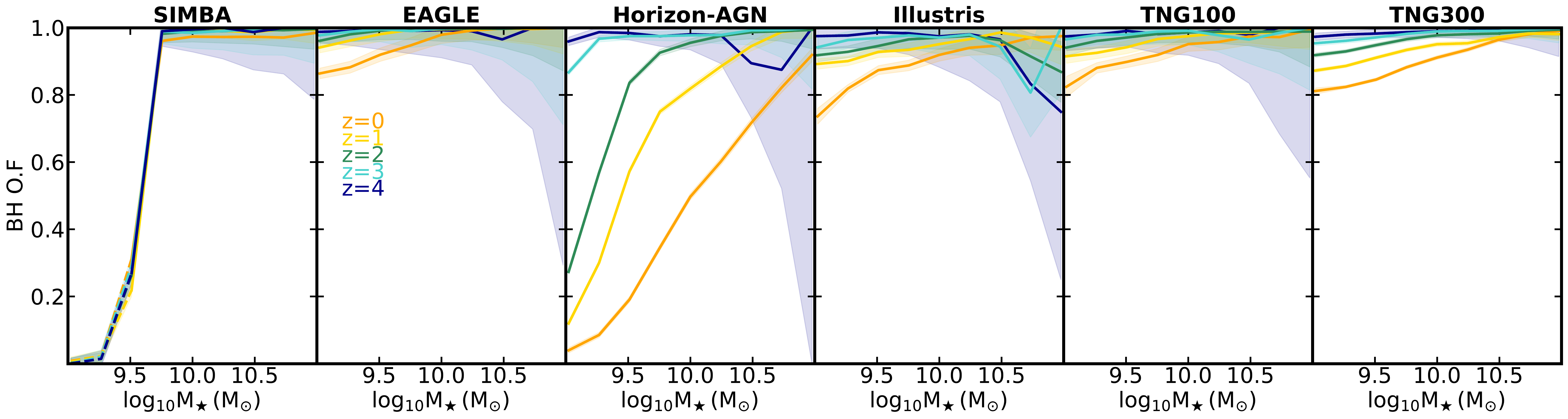}
    \caption{
    The BH occupation fraction at redshift $z=0-4$ predicted by the all six simulations, over the range $10^{9} \,{\rm M_{\odot}}\leqslant M_{\star}\leqslant 10^{11} \,{\rm M_{\odot}}$ (including both central and satellite galaxies). Shaded areas represent Poisson error bars. All the simulations have a higher BH occupation fraction at higher redshift, at fixed stellar mass. With time, the fraction drops in the low-mass galaxy regime. This is mainly due to a larger number of satellite galaxies that do not host a BH.The decrease is stronger in the Horizon-AGN simulation. }
    \label{fig:BHoccfrac-time-evo}
\end{figure*}

\subsection{Comparisons with current observational constraints in the local Universe}
In Fig.~\ref{fig:BHoccfrac-err}, we show the current observational constraints on the BH OF at $z=0$. Here, we only discuss the agreement of the simulations with the constraints. The implications and uncertainties are later reviewed in the Discussion section. 

One of these constraints comes from BH dynamical mass measurements 
performed for five of the nearest low-mass (with  
$M_{\star}=5\times 10^8-10^{10}\, \rm M_{\odot}$) early-type galaxies M32, NGC 205, 
NGC 5102, NGC 5206
\citep{2018ApJ...858..118N}, and NGC 404 
\citep[][]{2017ApJ...836..237N,2020MNRAS.496.4061D}.
While small in number statistics, this sample of five early-type galaxies (ETGs) is 
volume limited in the considered stellar mass range but only targets nucleated galaxies, and provides a BH occupation of $80\%$ for these galaxies.
The OF in ETGs was also investigated in \citet{2008ApJ...680..154G} with X-ray source detections in $16\%$ of their sample galaxies. The AGN detections were combined to the X-ray luminosity function to derive a constraint on the BH occupation 
from their ``clean''
sample ($1 \sigma$ confidence) for galaxies within the range $M_{\star}= 10^{7.5}-10^{12.5}\, \rm M_{\odot}$ 
\citep{2015ApJ...799...98M}.

We also consider the constraints from \citet{trump2015biases} (SDSS data), for which the BH OF is fitted to the AGN OF, assuming that the population of BHs in galaxies would be composed of 
low-mass BHs (possible formed via the PopIII remnant model) 
and more massive BHs 
(formed via heavy seed models).
These fits as a function of different galaxy quantities such as specific SFR (sSFR), galaxy mass-color, and  concentration, indicate that BHs in low-mass galaxies would likely be undermassive in the $M_{\rm BH}-\sigma$ relation. Their BH OF is derived from the high-mass regime of this relation. Their fits from 
more than 200 000 galaxies indicate that BHs would be present in about $30$ to $50\%$ of low-mass galaxies.

All simulations except Horizon-AGN produce 
a BH OF consistent with the observational constraint derived by \citet{2015ApJ...799...98M}, and are above the constraints derived for BH dynamical mass measurements of \citet{2017ApJ...836..237N,2018ApJ...858..118N}. 
Horizon-AGN has a BH OF that is 
below the constraints of \citet{2015ApJ...799...98M,2017ApJ...836..237N,2018ApJ...858..118N}.  
However, it is the only simulation to be in agreement with the constraints of \citet{trump2015biases}.

Finally, we also add the observational constraints of \citet{2019ApJ...878...18S,hoyer2021nucleation} \citep[see also][]{2021arXiv210503440C} on the presence of nuclear star clusters in hundreds of galaxies in the nearby Virgo galaxy cluster and the Local Volume $(d \leqslant 12\,\rm Mpc)$. While not directly constraining the BH OF, there is evidence for the presence of BHs within nuclear star clusters in the local Universe, while 
there exist also cases of nuclear star clusters devoid of BHs \citep{2020A&ARv..28....4N}. \citet{2019ApJ...878...18S} investigate the presence of nuclear clusters in hundreds of galaxies in the center of Virgo. The nucleation fraction reaches $\sim 80-90\%$ for galaxies with $M_{\star}\sim 10^{9}\, \rm M_{\odot}$, and decreases towards less and more massive galaxies. For the Local Volume, \citet{hoyer2021nucleation} find that the nucleation fraction reaches $80\%$ in galaxies with $\sim 10^{9}\, \rm M_{\odot}$, up to slightly more than $80\%$ in galaxies of $\sim 10^{10}\, \rm M_{\odot}$. We will discuss these constraints in Section~\ref{sec:NSC}.

\subsection{Evolution of the BH occupation fraction with redshift}
\label{sec:sec3.3}
Fig. \ref{fig:BHoccfrac-time-evo} shows that SIMBA achieves an OF of about unity at $\rm M_{\star} \geqslant 10^{9.5} M_{\odot} $, and stays that way throughout the redshift range $z=0-4$.
This is a direct result of the seeding model: all galaxies with $M_{\star}\geqslant 10^{9.5}\, \rm M_{\odot}$ are seeded with a BH at any redshift.
The other simulations have BH OF close to unity at high redshift ($z=4$). These fractions all decrease with decreasing redshift at fixed stellar mass for $M_{\star}\leqslant 10^{10.5}\, \rm M_{\odot}$, although with different amplitudes. 
Taking $z=2$ for example, the BH OF  at $\rm M_{\star} = 10^{9.5} M_{\odot}$ in EAGLE and Horizon-AGN are $97\%$ and $78\%$ respectively.

Eagle, Illustris and TNGs all have a halo mass threshold for the seeding of $\sim 10^{10}\, \rm M_{\odot}$. 
BH particles can spawn at any time if the condition for formation is fulfilled.
The reason for the decreasing trend with decreasing redshift is that some low-mass satellite galaxies can lose their BHs due to BH repositioning when they are close to their central galaxies. This artificial process is more common at lower redshift, and is responsible for the lower OF at lower redshift. 

The decrease of BH OF with decreasing redshift is the most pronounced in Horizon-AGN. This is a combination of several factors. 
 In this simulation, there is an exclusion radius to form new BHs: with time new low-mass galaxies form and if they are close to a large galaxy which already has a BH they will not form their own BHs.
Furthermore, no BH particles are formed after $z\sim 1.5$, therefore low-mass galaxies forming after $z\sim 1.5$ do not host a BH \citep[see also][]{2016MNRAS.460.2979V, 2017MNRAS.468.3935H}. Finally, seeding in Horizon-AGN takes place in cells whose local gas density is higher than the density to form stars, and as gas density decreases at low redshift, so does the probability of seeding.

\subsection{Simulated BHs in the dwarf galaxy regime}

\begin{figure*}
    \hspace*{-0.6cm}
    \includegraphics[scale=0.29]{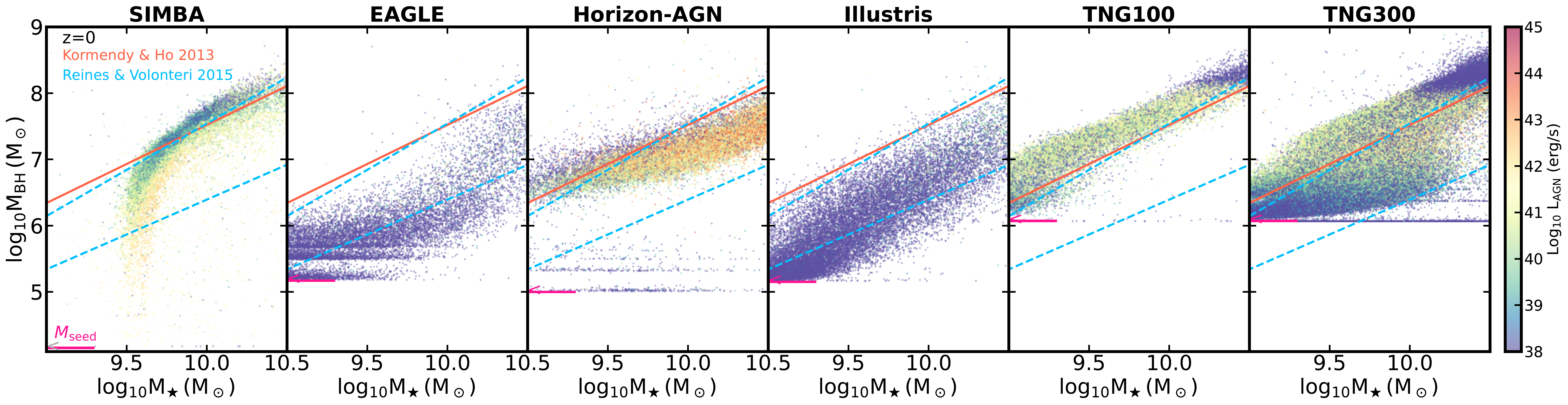}
    \caption{ 
    $\rm M_{BH}-M_{\star}$ diagram for galaxies with $M_{\star}=10^{9}-10^{10.5}\, \rm M_{\odot}$ at $z=0$, color coded as a function of the AGN hard X-ray luminosity (2-10 keV). 
    The pink arrows point towards the seeding mass of the simulations. 
    We overlay the $\rm M_{BH}-M_{\star}$ scaling relations from \citet{reines2015relations} and \citet{2013ARA&A..51..511K} as a reference.
    The different $ M_{BH}-M_{\star}$ diagrams of the simulations, and the diversity of AGN luminosities at fixed BH mass and galaxy total stellar mass indicate that simulations produce different AGN populations in the low-mass  galaxy regime. 
}
    \label{fig:mgal-mbh-scatter}
\end{figure*} 

\begin{figure*}
    \hspace*{-1cm}
    \includegraphics[scale=0.295]{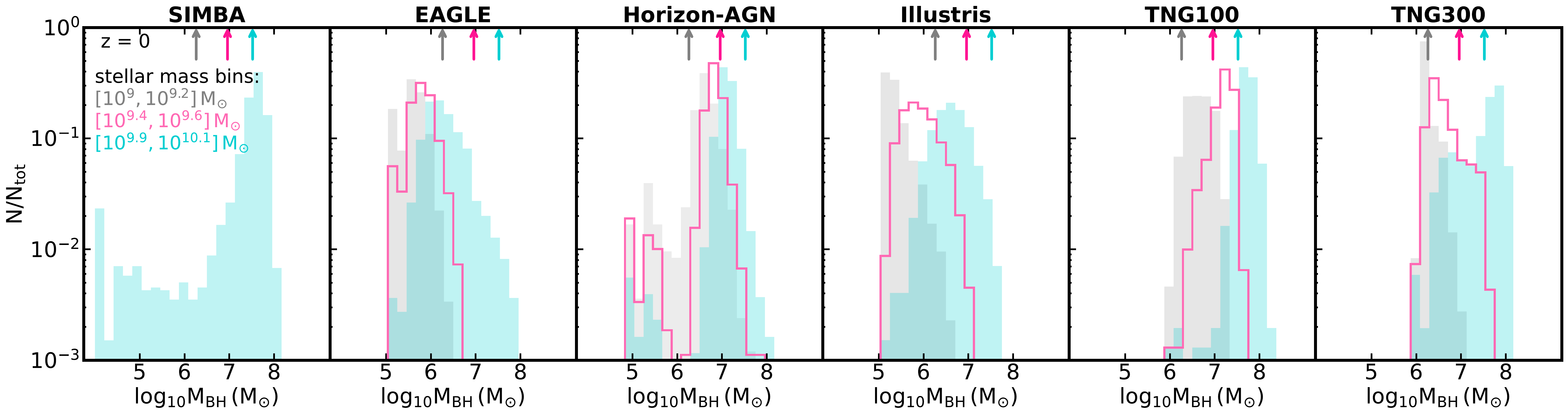}
    \caption{The normalised distribution of BH mass at $z=0$ in three  different stellar mass bins: $10^{9}\leqslant M_{\star}/\rm M_{\odot}\leqslant10^{9.2}, 10^{9.4} \leqslant M_{\star}/\rm M_{\odot} \leqslant 10^{9.6}$, and $10^{9.9}\leqslant M_{\star}/\rm M_{\odot}\leqslant10^{10.1}$.
    Arrows point to the mass of the most massive BHs in the observational sample of \citet{reines2015relations} derived in each of the three stellar bins. 
    For SIMBA, we do not display the BH population for $M_{\star} <10^{9.5}\, \rm M_{\odot}$ since the simulation starts seeding BHs at the aforementioned value. 
    The mass of BHs in low-mass galaxies strongly depends on the seeding mass in some simulations with strong SN feedback and/or low BH accretion rates, and does not in simulations with weak SN feedback, and/or efficient BH accretion, and/or accretion that does not strongly depend on BH mass. For all the stellar mass bins presented here, BH accretion is dominant over BH mergers.}
    \label{fig:mbh-hist}
\end{figure*}

While almost all the simulations produce a BH OF of $\geqslant 70\%$ in dwarf galaxies of $10^{9}\, \rm M_{\odot}$, the mass of the BHs that they contain vary from one simulation to another. This is due to a combination of the seeding mass, BH accretion and galaxy modeling \citep{2019arXiv191000017L,2021MNRAS.503.1940H}. 

To investigate the BH population produced by the different simulations, we resort to the $M_{\rm BH }-M_{\star}$ diagram, shown in Fig.~\ref{fig:mgal-mbh-scatter}. 
Most simulations show that the $M_{\rm BH}$ scatter is higher in the galaxy low-mass regime ($M_{\star}\leqslant 10^{10.5}\, \rm M_{\odot}$) and decreases with increasing mass of both the BH and the host galaxy \citep[$M_{\star}> 10^{10.5}\, \rm M_{\odot}$, shown in Fig.~ 2 of][]{2021MNRAS.503.1940H}. Besides having the same $M_{\rm BH}$ scatter trend with stellar mass, the simulations produce different BH mass ranges at fixed stellar mass. In the range $M_{\star}\leqslant 10^{10}\, \rm M_{\odot}$, BHs are on average more massive in Horizon-AGN, TNGs, and SIMBA, than they are in EAGLE and Illustris. These differences produce different shape of the $M_{\rm BH}-M_{\star}$ diagrams. Some simulations such as Horizon-AGN, Illustris, SIMBA and TNG100 (at low redshift) employ a modeling of SN feedback (either weak, or decoupled winds) that has a relatively weak effect on the growth of BHs. BHs can grow efficiently with their galaxies, resulting in a linear $M_{\rm BH}-M_{\star}$ relation.
However, a simulation such as EAGLE regulates more strongly the growth of BHs with respect to the host galaxies in the low-mass galaxy regime, resulting in a plateau at the low-mass end of the $M_{\rm BH}-M_{\star}$ diagram. This is due to a combination of stronger SN feedback and on average lower accretion rates due to the EAGLE modified Bondi model \citep{2016MNRAS.462..190R}.

To quantify the results presented above, we show in Fig.~\ref{fig:mbh-hist} the mass histograms of BHs embedded in galaxies with $M_{\star}\sim 10^{9}, 10^{9.5}, 10^{10}\, \rm M_{\odot}$ for all the simulations. The histograms are normalized by the number of BHs in each stellar mass bin. While EAGLE and Illustris have similar BH mass distributions for $ \rm M_{\star}\sim 10^{9}\, \rm M_{\odot}$, with BHs mostly in the range $ \rm M_{\rm BH}\sim 10^{5}-10^{6}\, \rm M_{\odot}$, Horizon-AGN and TNG  have BHs in the range $ \rm M_{\rm BH}\sim 10^{6}-10^{7}\, \rm M_{\odot}$.  
The Horizon-AGN BHs are able to grow efficiently from their seeding mass due to a relatively weak SN feedback, to produce $10^{6}-10^{7}\, \rm M_{\odot}$ BHs in dwarf galaxies. 
The high-mass BHs in galaxies of $10^{9-9.5}\, \rm M_{\odot}$ produced by some simulations appear to be more massive than the current observational constraints. In observations, BHs in these galaxies do not exceed $M_{\rm BH}\sim 10^{6.5}\,\rm M_{\odot}$ \citep{reines2015relations,2020ARA&A..58..257G,2020A&ARv..28....4N}. 

\begin{figure*}
    \includegraphics[scale=0.47]{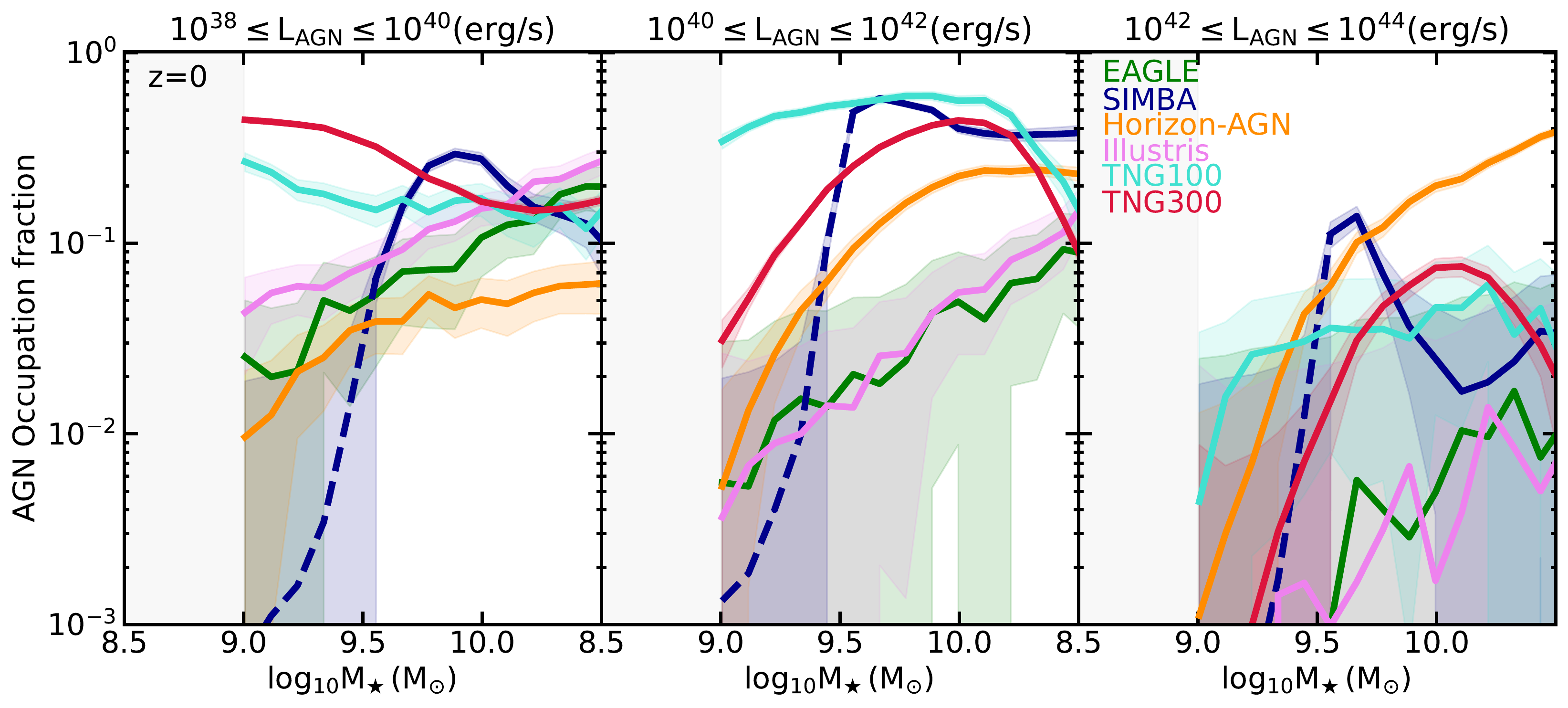}
    \includegraphics[scale=0.47]{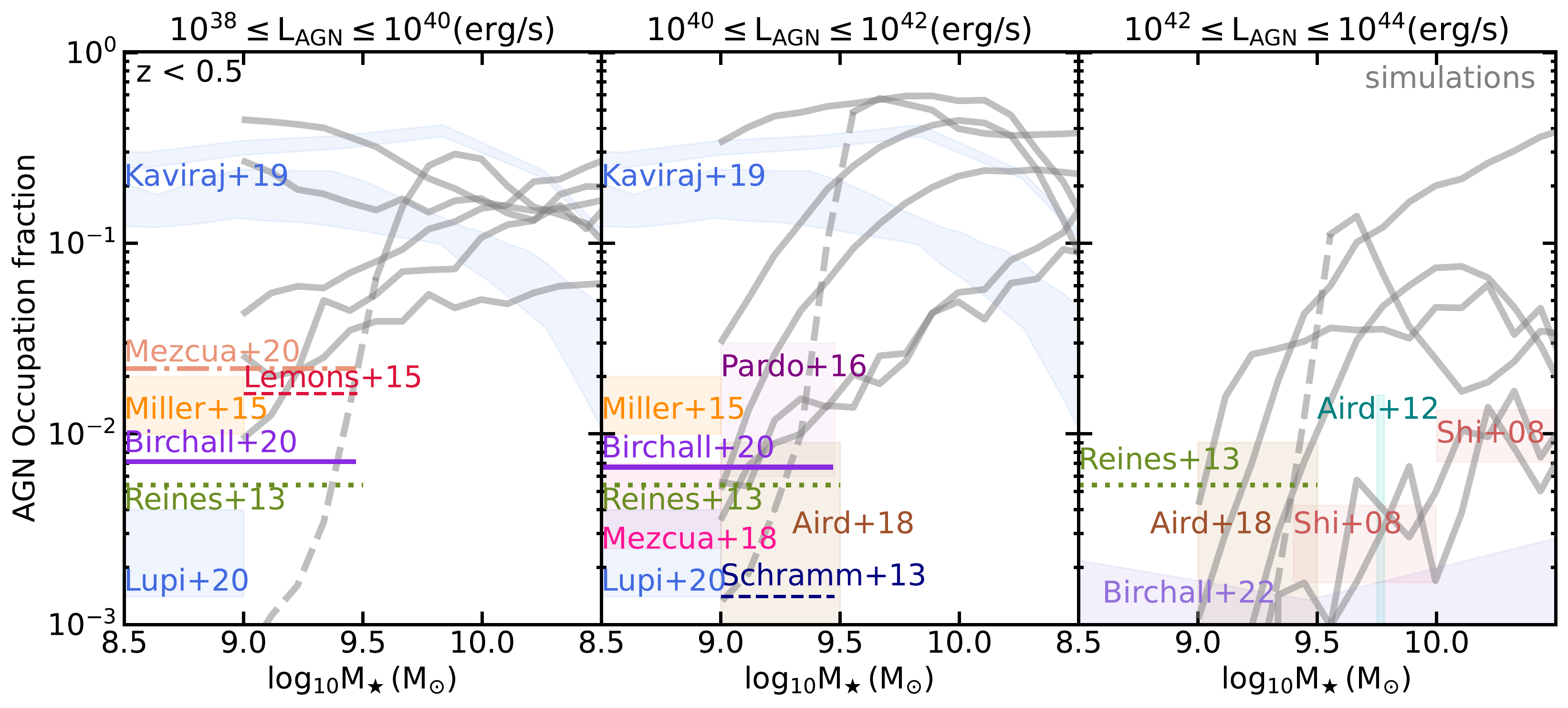}
    \caption{
    {\it Top figure}:
    AGN occupation fraction produced by the simulations in the low-mass regime ($10^{9} \leqslant M_{\star}/\rm M_{\odot}\leqslant 10^{10.5}$) at $z=0$. The different panels show three different  AGN luminosity ranges in hard X-ray (2-10 keV): $10^{38}\leqslant L_{\rm AGN}/( \rm erg/s) \leqslant 10^{40}$, $ 10^{40}\leqslant L_{\rm AGN}/ \rm (erg/s)\leqslant 10^{42}$ and $ 10^{42}\leqslant L_{\rm AGN}/(\rm erg/s) \leqslant 10^{44}$. The shaded areas represent Poisson errors.
    The AGN fractions in the simulations span several orders of magnitude.
    {\textit{Bottom figure}}: Current observational constraints on the AGN occupation fraction for low-mass galaxies ($ 10^{9.5} \leqslant  M_{\star}/\rm M_{\odot} \leqslant 10^{10.5}$) for $z<0.5$. We caution here that not all these constraints include a correction to account for incompleteness of the observational samples. 
    We overlay the AGN occupation fraction derived from the simulations, at $z=0$, in grey.
}
    \label{fig:AGN_obs_vs_sim}
\end{figure*}

The BH mass distribution in more massive galaxies with $  M_{\star}\sim 10^{10}\, \rm M_{\odot}$ also varies from one simulation to another. Such galaxies host BHs of $M_{\rm BH}\sim 10^{5}-10^{8}\, \rm M_{\odot}$ on average across simulations. 
In EAGLE and Illustris, BHs of $  M_{\rm BH}\sim 10^{5}-10^{8}\, \rm M_{\odot}$ are found in these galaxies, with distributions peaking at $M_{\rm BH}\sim 10^{6-7}\, \rm M_{\odot}$. Fig. \ref{fig:mbh-hist} shows that the distribution in Horizon-AGN is centered on massive BHs with $M_{\rm BH}\sim 10^{7}\, \rm M_{\odot}$, while even more massive BHs with $M_{\rm BH}\sim 10^{7}-10^{8}\, \rm M_{\odot}$ are found in SIMBA and TNG100. The BHs in SIMBA are able to grow efficiently from their seeding mass of $  M_{\rm BH}\sim 10^{4}\, \rm M_{\odot}$ (lowest of all simulations) to be the most massive in galaxies with $  M_{\star}\sim 10^{10}\, \rm M_{\odot}$, among all the simulations.
The $M_{\rm BH}-M_{\rm \star}$ relation in SIMBA is independent 
of seed mass, because gravitational torque accretion has a very weak dependence on $M_{\rm BH}$ \citep{2013ApJ...770....5A,2015ApJ...800..127A,2017MNRAS.464.2840A}. SIMBA allows BHs in the torque-limited accretion mode to accrete up to $3\times$ the Eddington limit, while the other simulations are capped at Eddington. This can have some effect on the early growth of BHs \citep{2019MNRAS.486.2827D}.

\section{Results: AGN occupation fraction}
\label{sec:4}
\label{sec:agn_occ_fraction}
Using the simulations as a laboratory to study the AGN OF in the low-mass galaxies may aid in revealing details on the BH population and their host galaxies. We define the AGN OF as the ratio between the number of AGN and the total number of galaxies in bins of galaxy stellar mass.

\subsection{Local Universe}
In Fig.~\ref{fig:AGN_obs_vs_sim} (top panels), we show the AGN OF for three different AGN  luminosity ranges in hard X-ray (2-10 keV): $L_{\rm AGN} = 10^{38}-10^{40}$, $10^{40}-10^{42}$ and $10^{42}-10^{44} \, \rm  erg/s$\footnote{The AGN luminosity distributions of the simulations at $z=0$ are shown in Appendix~\ref{sec:AGN_lum_distri} (Fig.~\ref{fig:Lxray-hist}).}. The shaded areas represent the uncertainties assuming a Poisson distribution. 
For a given simulation, the fraction depends on the number of galaxies hosting a BH (as described in the previous section), and on the ability of galaxies to feed the existing BHs. 
The BH OF of Horizon-AGN is low in low-mass galaxies, and this is reflected in the AGN fraction.

We first analyze the evolution of the AGN fractions with AGN luminosity.
Low-mass galaxies are on average more likely to host a faint AGN with $L_{\rm AGN}=10^{38}-10^{42}\, \rm erg/s$ (two left panels) than a bright one with $L_{\rm AGN}=10^{42}-10^{44}\, \rm erg/s$ (right panel). In general, the presence of faint AGN in dwarf galaxies is due to a combination of the low-mass BHs in these galaxies (Fig.~\ref{fig:mbh-hist}), but also of the effect of SN feedback preventing BH growth or the modeling of BH accretion (e.g., in EAGLE). 
We note a different behavior in TNG100, for which the AGN fractions are higher when considering AGN with $L_{\rm AGN}=10^{40}-10^{42}\, \rm erg/s$ than fainter or brighter AGN.
At $z=0$, the higher fraction of AGN with $L_{\rm AGN}\leqslant 10^{42}\, \rm erg/s$ in $10^{9}-10^{10}\, \rm M_{\odot}$ galaxies in TNG100 compared to the other simulations comes from the combination of the high seeding mass of $\sim 10^{6}\, \rm M_{\odot}$ and a relatively weaker SN feedback\footnote{The strength of SN feedback in TNG changes with metallicity, and weakens with decreasing redshift \citep{2018MNRAS.475..648P}.}. The high seeding mass favors higher accretion rates (if gas is available) in the Bondi formalism because $\dot{M}_{\rm BH}\propto M_{\rm BH}^{2}$. The BH accretion in SIMBA is governed by a gravitational torque limited model for accretion from cold gas $(T < 10^{5} \, \rm K)$, and Bondi model in the presence of hot gas \citep[$T > 10^{5} \, \rm K$,][]{2019MNRAS.486.2827D}. Relatively low-mass BHs of $M_{\rm BH}\sim 10^{6}\, \rm M_{\odot}$ have high Eddington ratios $\log_{10}\, f_{\rm Edd}$ in the range (-2,0)
\citep[Fig.~5 of][]{2019MNRAS.487.5764T}. The presence of cold gas in the host galaxies of these BHs, and their high gas fraction, provide the BHs with the additional torque-limited accretion channel.
This explains the relatively high AGN OF for AGN with $L_{\rm AGN}=10^{42}-10^{44}\, \rm erg/s$ located in $\sim 10^{9.5}\, \rm M_{\odot}$ galaxies.

Regardless of the luminosity range, 
Fig. \ref{fig:AGN_obs_vs_sim} (top panels) shows that the 
AGN OF evolves with stellar mass, and varies from $0.001$  
to $0.6$ 
(overall estimate derived from all the simulations together) when considering all AGN with hard X-ray luminosities above $10^{38}\, \rm erg/s$. 

Focusing on the stellar mass range $M_{\star} = 10^{9}-10^{10.5}\, \rm M_{\odot}$, more massive galaxies have in general a higher probability of hosting an AGN, for any AGN luminosity. This is not the case for the SIMBA and TNG simulations. In the case of the TNGs, this is explained by the efficient low-accretion kinetic mode of the AGN feedback taking place. The self-regulation makes the AGN weaker, shifting them to the $L_{\rm AGN}=10^{38}-10^{40}\, \rm erg/s$ panel. To conclude, all simulations predict a low intrinsic fraction of AGN with $10^{42} \leqslant L_{\rm AGN}/\, \rm (erg/s) \leqslant 10^{44}$ in local low-mass galaxies with $M_{\star}\leqslant 10^{10.5}\, \rm M_{\odot}$ (below $40\%$), while half of the simulations predict an even lower fraction (below $10\%$).

\subsection{Comparisons with observational constraints}

We now turn to compare the AGN occupation fractions in simulations and observations, shown in Fig.~\ref{fig:AGN_obs_vs_sim} (bottom panels). To do so, we first compile an exhaustive list of the existing constraints from the literature \citep{2008ApJ...688..794S,Aird2012,schramm2013black,reines2013dwarf,2015ApJ...799...98M,lemons2015x,pardo2016x,2018MNRAS.474.1225A,2018MNRAS.478.2576M,2019MNRAS.489L..12K,2020ApJ...898L..30M,2020MNRAS.492.2268B,2022MNRAS.510.4556B} that we report in Appendix~\ref{sec:observations_list}, along with a detailed discussion on how results are treated and homogenized.  We use a cut in hard X-ray luminosity to derive the occupation fractions in the simulations. In some cases, however, the observational constraints are not for the exact same AGN hard X-ray luminosity as what we employ for the simulations.  Other wavelengths 
and selection methods (e.g., emission line ratios, BPT diagrams) are sometimes used to identify AGN. 
The results of \citet{reines2013dwarf} are reported in all panels because they can not be connected to hard X-ray luminosities, and the results of \citet{2015ApJ...799...98M} in two panels because they can not be divided further. 
We sometimes extrapolate the results of some of these studies to derive an OF (when not already present in the respective papers).  Finally, only some of these works employ a correction for completeness \citep[often based on X-ray deep surveys,][]{Aird2012,2015ApJ...799...98M,pardo2016x,2018MNRAS.474.1225A,2018MNRAS.478.2576M,2020MNRAS.492.2268B}; this is a key aspect when computing an observational constraint on the AGN OF, and we detail this in Appendix~\ref{sec:observations_list}. 
While our comparison between the simulations and the observations presented in Fig.~\ref{fig:AGN_obs_vs_sim} carries uncertainties (because of e.g., observation completeness, observations in different wavelengths), it provides us with qualitative results.

Most constraints indicate an AGN OF in dwarf galaxies, across all luminosity ranges, of $\leqslant 0.03$, with significant variations over more than an order of magnitude. 
\citet{2019MNRAS.489L..12K} find an enhanced  presence of AGN in low-mass galaxies, with an AGN OF  $\rm \geqslant 0.01$ for relatively faint AGN. 
Detecting the presence of AGN in dwarf galaxies is particularly difficult. Star formation can be a major contaminant at many wavelengths and is also aperture dependent. 
Recently, \citet{2020MNRAS.492.2528L} carry out a second analysis of the same sample, considering contamination from close sources, and contamination from the stellar galaxy component, and they find a much lower AGN fraction of about $0.0014-0.004$.

For relatively bright AGN ($L_{\rm AGN}>10^{40}\, \rm erg/s$, top and middle panels of Fig.~\ref{fig:AGN_obs_vs_sim}), most simulations are in broad agreement with observations. TNG100, TNG300 and SIMBA have overall higher AGN OF.  
In SIMBA this enhanced activity is because BHs need to accrete efficiently to reach the $M_{\rm BH}-M_{\star}$ scaling relation, by design of the simulation. At the faint end, $L_{\rm AGN}=10^{38}-10^{40}\, \rm erg/s$,  
we find a good agreement between  EAGLE, Horizon-AGN  and observations.  Illustris and TNGs produce about one order of magnitude more faint AGN in dwarf galaxies
because of the relatively weak SN feedback employed at low redshift and/or the high seeding mass \citep{2018MNRAS.475..648P,2021MNRAS.503.1940H}. 

This analysis, however, carries large uncertainties 
from both the observational and theoretical side. In observations, completeness is not always stated in papers or easily inferred. In some simulations, BHs in low-mass galaxies are more massive  compared to observations (Fig.~\ref{fig:mgal-mbh-scatter}, Fig.~\ref{fig:mbh-hist}), and this could lead to more and/or brighter AGN, thus artificially enhancing the AGN OF. 
It is also important to mention that we did not correct the simulations for possible gas and dust obscuration of the AGN. The AGN OF derived from the simulations can thus be considered as upper limits when compared to observational constraints. We discuss this further in the Section \ref{sec:6}.

\begin{figure*}
    \centering
    \hspace*{-0.5cm}
    \includegraphics[scale=0.28]{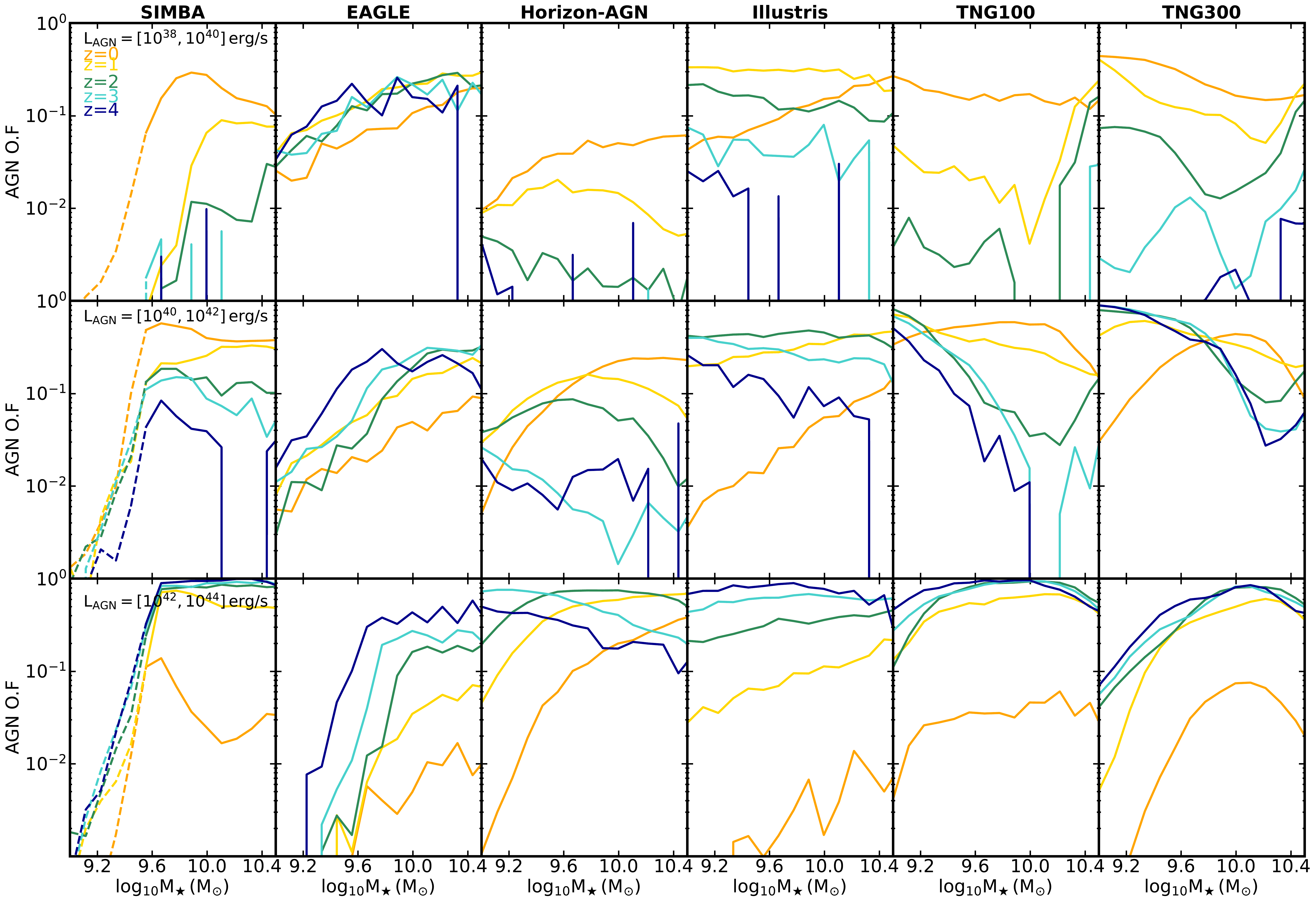}
    \caption{The time evolution of the AGN occupation fraction for $z=0-4$  in low-mass galaxies $10^{9} \leqslant M_{\star}/{\rm M_{\odot}} \leqslant 10^{10.5}$. AGN are selected on the basis of three different X-ray luminosity ranges: $  10^{38}\leqslant L_{\rm AGN}/\rm (erg/s) \leqslant 10^{40}$ (top row), $  10^{40}\leqslant L_{\rm AGN}/\rm (erg/s) \leqslant 10^{42}$ (middle row) and $  10^{42}\leqslant L_{\rm AGN}/\rm (erg/s) \leqslant 10^{44} $ (bottom row).
    For visibility, we do not display here the Poisson errors (but see Fig \ref{fig:AGN_obs_vs_sim} for an example at $z=0$).
}
    \label{fig:AGNOF-high-redshift}
\end{figure*}

\begin{figure*}
    \centering
    \includegraphics[scale=0.34]{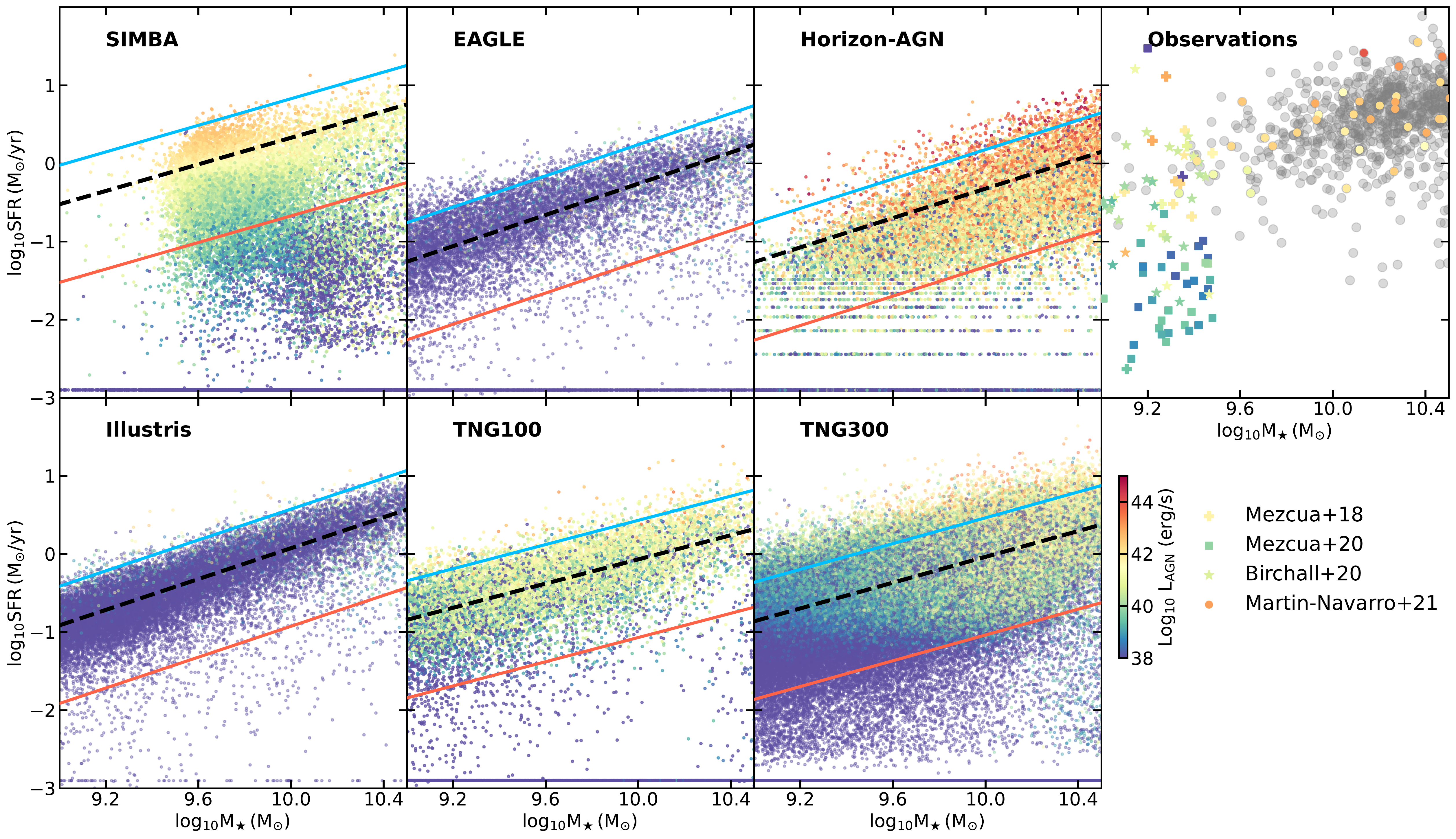}
    \caption{
    SFR-$M_{\star}$ diagram at $z=0$ for the six cosmological simulations. Galaxies with SFR lower than $10^{-3}\, \rm M_{\odot}/yr$ are set to this value. Galaxies are color coded by the hard X-ray luminosity of their AGN. We set all AGN luminosity values lower than $L_{\rm AGN} = 10^{38}\, \rm ergs/s$ (including inactive BHs)
    to the aforementioned value.
    We show the star-forming main-sequence of each simulation with a black dashed line. 
    Starburst galaxies are located above the blue lines, and quiescent galaxies below the red ones.
    In the right panel, we show several recent samples of galaxies from \citet{2020ApJ...898L..30M,2020MNRAS.492.2268B}, and \citet{2021MNRAS.tmpL..99M}. Observed galaxies are color coded  (when possible, otherwise in grey) according to the luminosity of their AGN, as for the simulations. 
    For a fair comparison, we have converted the AGN luminosity of several of these samples to hard (2-10 keV) X-ray luminosity.
    }
    \label{fig:sfr-mgal-scatter}
\end{figure*}

\subsection{Evolution of the AGN occupation fraction with redshift}

\subsubsection{Evolution in the simulations}
The AGN OF varies significantly with redshift 
for all the simulations, often over more than one order of magnitude. This is shown in Fig.~\ref{fig:AGNOF-high-redshift} for three AGN hard X-ray luminosity ranges.

In SIMBA, Horizon-AGN, and TNGs, the number of galaxies with $M_{\star}=10^{9.5}-10^{10.5}\, \rm M_{\odot}$ powering AGN with $L_{\rm AGN}=10^{38}-10^{40}$, and $10^{40}-10^{42}\, \rm erg/s$ (top and middle rows of Fig.~\ref{fig:AGNOF-high-redshift}) increases with time, from $z=4$ to $z=0$. As such, these AGN are more preponderant in the nearby Universe. The evolution with time in Illustris and EAGLE is not exactly the same: the fractions increase with time down to $z\sim 1$, but then decreases down to $z=0$.

The number of galaxies producing brighter AGN with $L_{\rm AGN}=10^{42}-10^{44}\, \rm erg/s$ decreases with time for all the simulations, and even more so in the redshift range $z=1-0$. The strong decrease of these AGN at $z\leqslant 1$  is due to cosmic starvation: the quantity of gas available to feed the BHs decreases with time. At $z\geqslant 1$, instead, a significant fraction of low-mass galaxies may be harboring a relatively bright AGN, according to these simulations. 
For example, the AGN fraction 
($L_{\rm AGN} \geqslant 10^{42}\, \rm erg/s$)
in low-mass galaxies increases by at least $\sim 30\%$ from $z=0$ to $z=1$, in all simulations except EAGLE for galaxies of $\leqslant 10^{9.5}\, \rm M_{\odot}$.
Pushing observations to $z=1$ could allow us to build tighter constraints on the BH OF by detecting more AGN in low-mass galaxies with $M_{\star}=10^{9}-10^{10.5}\, \rm M_{\odot}$.

\subsubsection{Comparison with observations}
In observations, whether the AGN OF increases with redshift is still unclear. This is partly due to observations being often biased towards more luminous AGN beyond the local Universe. \citet{2008ApJ...688..794S} find that the AGN OF slightly increases in $\sim 10^{10}\, \rm M_{\odot}$ galaxies from $0.1<z<0.4$ to $0.4<z<0.7$. \citet{Aird2012} also find that the AGN fraction, in fixed hard X-ray luminosity ranges (from $L_{\rm AGN}\sim 10^{42}\,\rm erg/s$ to $10^{44}\, \rm erg/s$), increases from $0.2<z<0.6$ to $0.6<z<1$ in galaxies with $M_{\star}\sim 10^{10.25}\, \rm M_{\odot}$.  Finally, \citep{2022MNRAS.510.4556B} also recently find a slight increase in the AGN fraction from $z=0$ to $z=0.3$ at fixed galaxy stellar mass (in the range $M_{\star}=10^{10-11}\, \rm M_{\odot}$).
We find the same trend in most of the simulations, a major difference being that 
the time evolution found in observations is smaller than the evolution in the simulations between $z=0$ and $z=1$.

On the other hand, several studies have found a decrease of the AGN fraction with redshift.
\citet{2018MNRAS.478.2576M} report a decrease for AGN with $L_{\rm x}\geqslant 10^{42.4}\, \rm erg/s$ in lower mass galaxies with $M_{\star}=10^{9}-10^{9.5}\, \rm M_{\odot}$ at $z\sim 0.4$, and a slight increase at $z\sim 0.6$ that is still lower than the AGN occupation they find at $z\sim 0.1-0.2$.
\citet{2018MNRAS.474.1225A} 
find a decrease of the AGN OF with redshift for AGN with $L_{\rm x}\leqslant 10^{42.4}\, \rm erg/s$ and an increase for brighter AGN.

As said in the previous section, the different subgrid physics employed in the simulations leads to large discrepancies in the AGN fractions at $z=1$, and to different evolution from $z=1$ to $z=0$. This makes $z=1$ a key regime to further investigate the AGN OF in the observations, to possibly help disentangle the simulation modeling.
The fact that observational studies have not reached yet a consensus on whether the OF increases towards higher redshift, also supports $z=1$ as a crucial regime to constrain.  
New X-ray surveys from high sensitivity missions such as Athena, or the AXIS and Lynx NASA concept missions, will lead to the discovery of fainter AGN beyond the local Universe. Getting tighter constraints on the AGN OF also requires to overcome detection limits for the host low-mass galaxies (e.g., with JWST given a wide enough survey), and to match these galaxies to the AGN surveys.

\section{Results: Properties of AGN host dwarf galaxies}
\label{sec:5}
Whether an AGN can be successfully detected does not only depend on its luminosity, but also on the properties of its host galaxy. 
For example, starburst activity in dwarf galaxies can be responsible for a high IR emission making IR detections of AGN challenging \citep[e.g.,][]{2020MNRAS.492.2528L}, but is also responsible for X-ray emission from binaries which can contaminate the AGN emission \citep{2018MNRAS.478.2576M}. 
To understand how many simulated AGN could be detected in observations, we investigate the mass and SFR properties of the AGN host galaxies.  

\subsection{Definition of the star-forming main-sequence}
\label{sec:5.1}

We start by computing the star-forming main sequence (MS, black dashed line in Fig.~\ref{fig:sfr-mgal-scatter}) for all the simulations at $z=0$,
by fitting their main SFR-$M_{\star}$ relation in the stellar mass range $M_{\star}=10^{9}-10^{10}\, \rm M_{\odot}$ that is not affected strongly by AGN feedback quenching. The star-forming main-sequence is defined as a power law:
\begin{eqnarray}
\log_{10} \frac{{\rm SFR}_{\rm MS}}{\rm M_{\odot}/yr}=\alpha + \beta \log_{10} \left( \frac{M_{\star}}{10^{10}\, \rm M_{\odot}}\right).
\label{eq:SFR-power-law}
\end{eqnarray}

Our definition is redshift and sample dependent, which is convenient since  the simulations do not produce the exact same MS. This also allows one to compare with observational samples in a consistent way. 
For SIMBA, EAGLE, Horizon-AGN, Illustris, TNG100 and TNG300, we find the slope $\beta= 0.85,0.99,0.94,0.98,0.77,0.82$  
and the intercept $\alpha = 0.32,-0.25,-0.32,0.08,-0.07,-0.04$, respectively. 
We define starburst galaxies as having SFR larger than $0.5 \, \rm dex$ above the MS (above the blue line in Fig.~\ref{fig:sfr-mgal-scatter}), and quenched galaxies as galaxies 1 dex below the MS\footnote{The definition of starburst and quenched galaxies in a given simulation is with respect to the MS of this simulation. 
Thus, a starburst galaxy in a given simulation may not be a starburst in another simulation (even with the same SFR).} (below the red line).
At $z=0$ for $\sim 10^{9}\, \rm M_{\odot}$ galaxies, the portion of starburst galaxies is about $5\%$ and that of quenched galaxies between $12\%$ and $ 33\%$.
The large fraction of low-mass galaxies with reduced SFR mostly results from the quenching of satellite galaxies \citep{2021MNRAS.500.4004D,2021ApJ...915...53D}. 
Main-sequence galaxies are the most predominant 
with values ranging between $61-86\%$ across the simulations.

\begin{figure*}
    \centering
    \hspace*{-0.2cm}
    \includegraphics[scale=0.38]{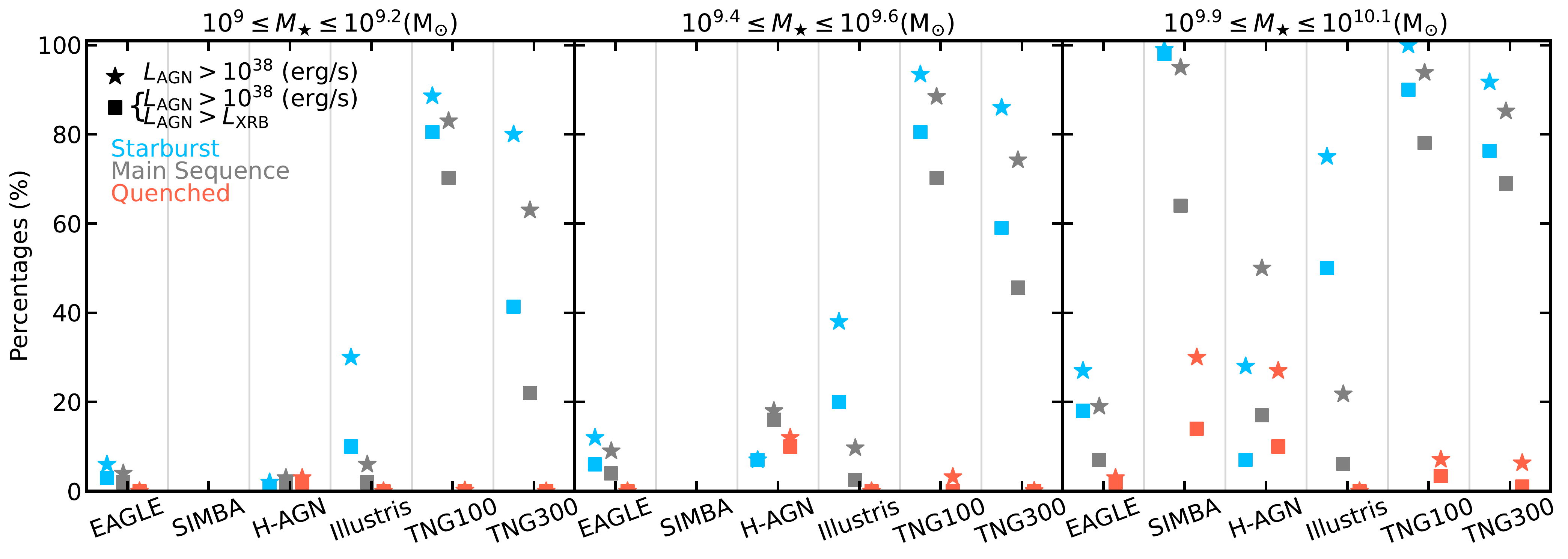}
    \caption{
    Percentages of galaxies at $z=0$ with hard X-ray luminosity $L_{\rm AGN} > 10^{38}\,\rm (erg/s)$ (shown as stars) and percentages of those galaxies which have a brighter AGN than the galaxy-wide XRB population, i.e., $L_{\rm AGN} > L_{\rm XRB}$ (shown as squares). We derive these percentages for different galaxy types color coded as: starburst (blue), main sequence (grey) and quenched (red). 
    While only a small fraction of quenched galaxies host an AGN in most of the simulations, the fraction among starburst and main-sequence is more important. Starburst galaxies are not always the galaxies hosting the largest number of AGN (see Horizon-AGN). The fraction of galaxies for which the AGN outshine the galaxy-wide XRB population is significantly lower for all galaxy types and simulations.
    }
    \label{fig:table2_visualisation}
\end{figure*}

\subsection{Star-forming properties of low-mass galaxies hosting AGN}
In Fig.~\ref{fig:sfr-mgal-scatter}, we show the SFR-$M_{\star}$ plane of the simulations. Each galaxy is 
color coded by its AGN hard X-ray luminosity.
The correlation between AGN luminosity and the location of galaxies in the SFR-$M_{\star}$ diagram vary from one simulation to another.
We find that, on average, brighter AGN are located in galaxies with higher SFR for all simulations at $z=0$. The trend with SFR is clearly visible in SIMBA, Horizon-AGN and TNGs. In these simulations, a galaxy actively forming new stars is also on average feeding its central BH efficiently. 
Such a positive correlation can be expected due to a common gas supply for star formation and BH accretion (particularly for some simulation subgrid physics), but high time variability of BH accretion rate and its small spatial scales complicates a direct one-to-one connection to global star formation 
\citep[e.g.,][]{hickox2014black,2021ApJ...917...53A}. 
EAGLE and Illustris produce populations of faint AGN and inactive BHs, which makes any trend between SFR, $M_{\star}$, and $L_{\rm AGN}$ difficult to identify. This is due to i) the fact that BH accretion rates are computed from the BH gas cells in Illustris, which makes the rates stochastic and not well correlated with the global SFR of the host galaxies, ii) the lower accretion rates produced by the accretion modeling of EAGLE compared to other simulations, and its efficient SN feedback.

We now turn to quantify the number of galaxies\footnote{We do not differentiate between isolated and non-isolated galaxies, nor between central and satellite galaxies.} 
of different types hosting an AGN. We show the percentages of starburst, main-sequence, and quiescent galaxies hosting an AGN with $L_{\rm AGN}\geqslant 10^{38}\, \rm erg/s$ in Fig.~\ref{fig:table2_visualisation} (star symbols). We report the percentages for $L_{\rm AGN}\geqslant 10^{38}$ and $\geqslant 10^{40}\, \rm erg/s$ in Table~\ref{tab:table-percentage-galtype-AGN}.

Most of the simulations do not produce any AGN (of $L_{\rm AGN} \geqslant 10^{38} {\rm or}\,10^{40}\, \rm erg/s$) in quenched galaxies with $M_{\star}\sim 10^{9}\, \rm M_{\odot}$. 
Discrepancies among the simulations arise in main-sequence galaxies. More than $50\%$ 
of the TNG100 main-sequence galaxies have an AGN with $L_{\rm AGN} \geqslant 10^{40}\, \rm erg/s$, while this number drops below $5\%$
for all the other simulations.
We identify the same trends for more massive galaxies with $M_{\star}\sim 10^{9.5}\, \rm M_{\odot}$ and $M_{\star}\sim 10^{10}\, \rm M_{\odot}$, although with on average slightly higher fraction of AGN.

Interestingly, starburst galaxies are not always the ones powering more AGN with $L_{\rm AGN} \geqslant 10^{40}\, \rm erg/s$. For example, in Horizon-AGN main-sequence galaxies are on average more successfully feeding their AGN. Horizon-AGN is also the only simulation producing  AGN in more than $10\%$ of $\sim 10^{10}\, \rm M_{\odot}$ quenched galaxies.  

Overall, there is no consensus in the simulations presented here on the fraction of AGN embedded in different types of galaxies based on their SFR. While a simulation like Horizon-AGN is able to power AGN in starburst, main-sequence, and a few quiescent galaxies, another simulation such as EAGLE will only power very few AGN in the main-sequence and starburst galaxies, and none in quiescent galaxies.

\subsection{Comparison with observations}

In the top right panel of Fig. \ref{fig:sfr-mgal-scatter}, we compile some of the most recent observations in nearby or low-redshift low-mass galaxies.
Because not all samples were obtained following the same observational techniques and selections, our study aims at qualitatively comparing in which low-mass galaxies AGN are powered both in observations and simulations.

In Fig.~\ref{fig:sfr-mgal-scatter}, we present the AGN samples of \citet{2018MNRAS.478.2576M,2020ApJ...898L..30M,2020MNRAS.492.2268B}. We also searched for X-ray counterparts to the AGN of \citet{2021MNRAS.tmpL..99M}.
Technical details of these samples, and how we extract the hard X-ray luminosity of the AGN, are presented in Appendix~\ref{sec:obs_SFR_Mgal}. 
When taken together, the observational samples  
show that on average galaxies with higher SFR host brighter AGN in $\sim 10^{9.5}\, \rm M_{\odot}$ galaxies, a feature that we also find in SIMBA, Horizon-AGN, and TNGs. The luminosities of the AGN are also in agreement with those produced in these simulations.

Galaxies hosting the brightest AGN in observations can have high SFR ($\rm \log_{10} SFR/( M_{\odot}/yr) \geqslant 0$ for $M_{\star}= 10^{9}-10^{10.5}\, \rm M_{\odot}$), with the most star-forming galaxies being absent from all the simulations studied here.  Conversely, the lowest mass, least star-forming galaxies host on average very faint AGN, although some can reach $L_{\rm AGN}\sim 10^{42}\, \rm erg/s$. For galaxies with $M_{\star} < 10^{9.6}\, \rm M_{\odot}$, when using the TNG definition of quenched galaxies, we find that in observations \citep{2020MNRAS.492.2268B,2020ApJ...898L..30M,2018MNRAS.478.2576M}
the brightest AGN in quenched galaxies have luminosities $L_{\rm AGN}< 10^{42.4}\, \rm erg/s$.
Almost no simulation shows AGN in quenched dwarf galaxies in the same mass range. SIMBA and Horizon-AGN, however, defy this with e.g., a $\sim 3-12\%$ probability of finding very faint AGN with $L_{\rm AGN} \geqslant 10^{38} \, \rm erg/s$ in the galaxy mass range $M_{\star}=10^{9}-10^{9.4}\, \rm M_{\odot}$. Although not shown in Fig.~\ref{fig:sfr-mgal-scatter},
\citet{dickey2019agn} suggest that it is possible to find AGN in quenched galaxies\footnote{\citet{dickey2019agn} use a definition of quenched galaxies different from ours, and based on 
the $H\alpha$ equivalent width 
and the Dn(4000) index.
}. Their work identifies 16 potential AGN hosts, out of a sample of 20 quenched galaxies with $M_{\star} = 10^{8.5}-10^{9.5}\, \rm M_{\odot}$.

\subsection{XRBs and AGN}
The choice of hard X-ray emission is favored in the search of AGN because, in this band, obscuration is significantly reduced. 
However, AGN in low-mass galaxies face the challenge of being outshined by XRBs, especially in star-forming galaxies where HMXBs can be numerous \citep[e.g.,][]{2019ApJS..243....3L}. 
In the following, 
we quantify the number of galaxies in which XRBs could prevent the detection of AGN. 
To that purpose, Fig.~\ref{fig:table2_visualisation} (square symbols, see also Table~\ref{tab:table-percentage-galtype-AGN}) shows the fractions of different type galaxies with $L_{\rm AGN}>L_{\rm XRB}$ for all the simulations.

We find that AGN with $L_{\rm AGN}\geqslant 10^{40}\, \rm erg/s$ do not suffer significantly from XRB contamination in all the simulations.
However, the fraction of galaxies with an AGN of $L_{\rm AGN}\geqslant 10^{38}\, \rm erg/s$ brighter than the galaxy-wide XRB luminosity ($L_{\rm AGN}> L_{\rm XRB}$) varies strongly from one simulation to another, for all the galaxy stellar mass bins and galaxy SFR types. For starburst galaxies in some simulations, only about half of the AGN with $L_{\rm AGN}\geqslant 10^{38}\, \rm erg/s$ are brighter than the host galaxy XRB population (SIMBA, Horizon-AGN), 
while in other simulations (e.g., SIMBA for galaxies with $\sim 10^{10}\, \rm M_{\odot}$ or TNG100) most of the AGN are brighter than the XRBs (i.e., close star and square symbols in Fig.~\ref{fig:table2_visualisation}). This depends on whether the simulations produce a global faint (e.g., EAGLE) or bright population of AGN (e.g., TNG100), and on the correlation between AGN luminosity and SFR. On average, the percentage of AGN with $L_{\rm AGN}> L_{\rm XRB}$ among the $L_{\rm AGN}\geqslant 10^{38}\, \rm erg/s$ AGN decreases from starburst to main-sequence, and to quiescent galaxies, for all galaxy masses. This is a direct consequence of AGN being generally fainter in these galaxies than in starburst galaxies.

We further explore the $z=1$ regime. 
We find that the fraction of AGN able to outshine the galaxy-wide XRBs population increases by $\sim10-30\%$ across the simulations. 
This is not true for EAGLE where the probability of finding an AGN brighter than an XRB at $z=1$ is still relatively the same as that at $z=0$.
According to most of the simulations, the detections of AGN in X-ray could be less challenged by XRBs at $z=1$ than at $z=0$. The limitation would come from our instrumental power to detect relatively faint AGN beyond the local Universe.

\section{Discussion}
\label{sec:6}
\label{sec:discussion}
\subsection{Impact of the subgrid physics}
Most cosmological simulations do not follow the current theoretical models of BH formation (e.g., PopIII remnant, compact stellar cluster, and direct collapse models), and instead seed massive halos or galaxies with a fixed mass BH. 
The seeding of some simulations leads to the presence of BHs exceeding $10^{6.5}\, \rm M_{\odot}$ in galaxies with stellar mass of $10^{9}-10^{9.5}\, \rm M_{\odot}$, while less massive BHs are found in observations. 
The decrease of the BH OF with time in almost all the simulations is mostly due to satellite galaxies loosing their BHs, and as such it 
is not a direct outcome of BH formation efficiency at high redshift.
This means that the broad agreement at $z=0$ between the simulations and observations in low-mass galaxies does not validate the seeding models employed in the large-scale cosmological simulations.

These large-scale simulations only form one BH per galaxy or halo, and some of them also re-position BH particles to their host galaxy's centers based on different criteria. 
BH dynamics is thus not followed on small scales, which can significantly affect the OF and location of low-mass BHs in galaxies \citep[e.g.][]{2021MNRAS.508.1973M}, and BHs merge as soon as they are within the resolution distance of $\sim 1\, \rm kpc$ following galaxy mergers. In most of the simulations, these aspects result in the presence of a single BH per galaxy and the absence of off-center BHs, despite the latter being observed in local dwarf galaxies \citep{2020ApJ...888...36R, 2020ApJ...898L..30M}. BHs are more likely to accrete efficiently if located in the potential well of their host galaxies due to being embedded in their central gas reservoirs.
Most simulations therefore likely provide an {\it optimistic} view on BHs' ability to accrete gas in dwarf galaxies.

Moreover, BH accretion in large-scale cosmological simulations is an unresolved physical process, and is modeled as subgrid physics.
In most of the simulations (except SIMBA), the accretion rate onto a BH is always proportional to $M_{\rm BH}^{2}$ (Bondi formalism). At fixed gas reservoir and fixed galaxy subgrid modeling, a more massive BH will accrete more than a less massive one. 
Since the BH masses are higher than in current observations in several simulations (Horizon-AGN, TNGs, SIMBA), then their AGN luminosities could be enhanced with respect to observations. It could explain the higher AGN luminosity function identified in \citet[][their Fig. 5, top panels]{2022MNRAS.511.3751H} for AGN with $L_{\rm} \leqslant 10^{43.5}\, \rm erg/s$ for the TNG simulations, and with $L_{\rm} \geqslant 10^{43.5}\, \rm erg/s$ for Horizon-AGN.
 An agreement with the observed AGN OF would imply that observed and simulated BHs have very different accretion properties: to reach the same luminosity observed BHs (which are lighter than simulated BHs) must have higher accretion rates. 
Depending on the resolution of the simulations, and how the accretion is computed (including the possible addition of a boost factor), the rates can be both under or over-estimated \citep{2017MNRAS.467.3475N,2021ApJ...917...53A}.

TNG100 produces the highest number of AGN (with $L_{\rm AGN}\geqslant 10^{40}\, \rm erg/s$) in low-mass galaxies of all the simulations, and this is mainly due to the high seed mass of $10^{6}\, \rm M_{\odot}$ and a less effective SN feedback at low redshift. 
In order to reduce the AGN OF below $0.03$, in line with the bulk of observational constraints, about $90\%$ of the TNG100 AGN should be obscured and missed in observations of dwarf galaxies. Alternatively, a large fraction of the AGN should not be located in high gas density regions and be off-centred, so that accretion is reduced. From the numerical point of view, more efficient SN feedback would also reduce the AGN fraction. 
On the other hand, good agreement between the simulations and the current constraints would mean that no AGN is obscured in low-mass galaxies, or that on average too few BHs are powering AGN in the simulation.

In this paper, we used the same radiative efficiency of 0.1 for all the simulations. Using a radiative efficiency of 0.2 (as employed for the BH accretion and AGN feedback model in the Illustris and TNG simulations) on average predicts a higher AGN fraction for $L_{\rm AGN}\geqslant 10^{40}\, \rm erg/s$ (Fig.~\ref{fig:AGN_obs_vs_sim}), resulting in more discrepancies between observations and simulations within that luminosity range.

\subsection{Impact of the simulation resolution}

The resolution of the simulations affect both the properties of the BHs and AGN, and consequently the AGN OF. 
One way to assess the impact of the resolution is comparing the sibling TNG simulations, whose subgrid model parameters are fixed and do not depend on the simulation resolution \citep{2018MNRAS.473.4077P}. 
The less resolved gas content surrounding the BHs in TNG300 leads to lower accretion rates, partly due to a more impactful SN feedback (particularly at high redshift, and in low-mass galaxies). 
This can be seen in Fig.~\ref{fig:mgal-mbh-scatter} and Fig.~\ref{fig:mbh-hist}, where the $M_{\rm BH}$ scatter at fixed stellar mass is larger in TNG300 than TNG100, due to more low-mass BHs.  
These more numerous low-mass BHs do not accrete gas efficiently in TNG300, as shown by the AGN luminosity color code in Fig.~\ref{fig:mgal-mbh-scatter}.
This reflects on the AGN OF:  TNG300 has fewer bright AGN $\geqslant 10^{40} \, \rm erg/s$, whereas TNG100 more faint AGN (see Fig.~\ref{fig:AGN_obs_vs_sim}).

\subsection{Faint AGN and AGN obscuration}
For relatively faint AGN in dwarf galaxies, observations are best conducted in galaxies with low SFR, for which the contamination from HMXBs should be smaller (as HMXB are short-lived). As we demonstrated here, the contamination from XRBs starts being important mostly for faint AGN with $L_{\rm x}\leqslant 10^{40}\, \rm erg/s$. Those AGN are powered by low-mass BHs in the simulations.
In this paper, we have not modeled the obscuration of the AGN when computing their X-ray luminosities. 
As X-ray photons penetrate more easily through the gas content, X-ray detections are often used for AGN and are less biased toward more luminous AGN. \citet{chen2017x} report the analysis of a sample of 10 low-mass AGN. In the dwarf regime ($M_{\star}\leqslant 3\times10^{9}\, \rm M_{\odot}$), the authors find that $20\%$ of the AGN do not show AGN-like optical narrow emission lines, and thus could be obscured in these wavelengths. 
Still low-mass galaxies could harbor a large fraction of heavily obscured AGN in X-ray, as hinted by X-ray stacking studied beyond the local Universe  \citep[][]{xue2012tracking,2016ApJ...817...20M}. 
AGN obscuration in a large fraction of low-mass galaxies could impact the comparison between simulations and observations.

\subsection{Nuclear star clusters and BHs in low-mass galaxies}
\label{sec:NSC}

In Fig.~\ref{fig:BHoccfrac-err}, we show observational constraints from \citet{2019ApJ...878...18S,hoyer2021nucleation} on the nucleation fraction in nearby galaxies.
While observing inactive BHs or faint AGN in local low-mass galaxies is extremely difficult, detecting nuclear star clusters in galaxies is possible today from the Local Group  $(d \leqslant 3 \,\rm Mpc)$ up to the Coma galaxy cluster ($\sim 100$ Mpc).

The presence of nuclear star clusters in galaxies can facilitate the detection and characterization of the possible central low-mass BHs through dynamical measurements \citep[e.g., enhanced velocity dispersion,][]{2018ApJ...858..118N}. A nuclear cluster can also theoretically enhance the number of tidal disruption events, up to a couple of orders of magnitude for galaxies with $M_{\star}\leqslant 3\times 10^{10}\, \rm M_{\odot}$ \citep{2020MNRAS.497.2276P}.
Nucleated galaxies are thus good locations to investigate the presence of massive BHs.
A few cases of the co-existence of BHs and nuclear clusters have been observed besides the Milky Way (see Table 3 in \cite{2020A&ARv..28....4N} and references therein). 
Whether BHs and nuclear clusters co-exist in the center of all local low-mass galaxies or only a fraction of them is still unclear. This currently prevents the extrapolation from the constraints on the nucleation fraction to a constraint on the BH OF.

In nearby nucleated galaxies, BHs tend to be overmassive compared to the nuclear clusters in galaxies with $M_{\star}\geqslant 10^{10}\, \rm M_{\odot}$ and undermassive in lower-mass galaxies as recently compiled in  
\citet[][their Fig. 13]{2020A&ARv..28....4N}. 
In galaxies with $M_{\star}\leqslant 10^{9}\, \rm M_{\odot}$, the stellar population of nuclear clusters is commonly old and metal poor. The migration of globular clusters (usually with old stellar populations) towards the center of galaxies could explain such metallicity and stellar ages.
The star clusters in more massive galaxies with $M_{\star}\geqslant 10^{9}\, \rm M_{\odot}$ have younger stellar populations, likely due to in-situ star formation.
In observations, the nucleation fraction is generally found to be decreasing for more massive galaxies.
For these aforementioned galaxies, the presence of a BH could disrupt the nuclear clusters (if present) and explain their absence  \citep{antonini2015coevolution}.

Understanding the possible co-existence of BHs and nuclear star clusters in cosmological simulations would require higher spatial resolution in the center of galaxies; there have been developments in this direction recently \citep{2019MNRAS.483.3488B,2021ApJ...917...53A}. This would likely not be enough as some theories of nuclear cluster formation involve the migration of globular clusters to galaxy centers. In that case, high resolution in the entire galaxies would be needed to capture the assembly of the clusters; this is currently not possible with cosmological simulations of $\sim 100\,\rm cMpc$ side length.

\section{Conclusion}
\label{sec:7}
In this paper, we performed a systematic analysis of the BH and AGN population in dwarf and low-mass galaxies with 
$M_{\star}=10^{9}-10^{10.5}\, \rm M_{\odot}$ in six large-scale cosmological simulations of volume $(\geqslant 100\, \rm cMpc)^{3}$.
We summarize our findings below.

\begin{itemize}
    \item The modeling of some simulations (seeding and/or galaxy physics) produces a population of BHs too massive in low-mass galaxies with respect to current local observations. For example, the BHs of $\geqslant 10^{6.5}\, \rm M_{\odot}$ in simulated galaxies with $\leqslant 10^{9.5}\, \rm M_{\odot}$   (Fig.~\ref{fig:mgal-mbh-scatter}, Fig.~\ref{fig:mbh-hist}).

    \item In simulations, the seeding of BHs in massive halos of $M_{\rm h} \sim 10^{10}\, \rm M_{\odot}$ or galaxies of $M_{\star} \geqslant 10^{9.5}\, \rm M_{\odot}$, or based on local gas properties, 
    leads to a BH OF of almost unity in low-mass galaxies at high redshift ($z\geqslant 3$, Fig.~\ref{fig:BHoccfrac-time-evo}). 

    \item At fixed stellar mass, the OF decreases with time in all simulations except in SIMBA (Fig.~\ref{fig:BHoccfrac-time-evo}). The decrease is driven by a larger number of satellite galaxies artificially losing their BHs. 
    As such, the BH OF at $z=0$ is not inherited from the intrinsic OF at high redshift. Comparison with observations at $z=0$ thus do not inform us on the accuracy of the seeding subgrid models.

    \item At $z=0$, almost all the simulations have a high BH OF in the range $70-90\%$ in galaxies with $M_{\star}\sim 10^{9}\, \rm M_{\odot}$ (Fig.~\ref{fig:BHoccfrac-err}). This is in broad agreement with some large observational constraints (coming from BH dynamical mass measurement and from extrapolation from AGN samples). Horizon-AGN has an OF of $\sim 10\%$, also in broad agreement with some constraints.
    
    \item The AGN OF in simulations spans from less than an order of magnitude in galaxies with $M_{\star}\sim 10^{10}\, \rm M_{\odot}$ to two (or more) orders of magnitude in galaxies with $M_{\star}\sim 10^{9}\, \rm M_{\odot}$ (Fig.~\ref{fig:AGN_obs_vs_sim}). 
    \item Current low-redshift observational constraints on the AGN OF span more than an order of magnitude (see compilation in Appendix~\ref{sec:observations_list}, Fig.~\ref{fig:AGN_obs_vs_sim}). Compared to observations, some simulations power too many AGN in low-mass galaxies, 
    except if e.g. almost all AGN in dwarfs are obscured.
    This could be due to a more massive BH population in low-mass galaxies in these simulations (Fig.~\ref{fig:mgal-mbh-scatter}, Fig.~\ref{fig:mbh-hist}).

    \item There is no consensus among the simulations on whether the fraction of AGN in low-mass galaxies is higher at higher redshift (e.g., $z\geqslant1$).  
    If so, it could allow us to put tighter constraints on the BH OF by observing more AGN (thus reducing the extrapolation from the AGN OF to BH OF). The combination of new X-ray facilities such as Athena, and new galaxy surveys (e.g., JWST) will be pivotal.

    \item The correlation between AGN luminosity and SFR found in observations of low-mass galaxies is also 
    produced in some simulations, but not all.
    At fixed galaxy stellar mass, starburst galaxies often 
    host more and brighter AGN than main-sequence and quiescent galaxies.
    This does not help in 
    constraining the AGN OF, as emission from star-forming regions prevent the AGN detections in several wavelength bands.
 
    \item We find that 
    AGN with $L_{\rm AGN}\geqslant 10^{40}\, \rm erg/s$ almost always outshine the galaxy-wide XRB population (in starburst, main-sequence, and quiescent galaxies of all the simulations). 
    As our ability to detect fainter X-ray AGN with $L_{\rm AGN}\geqslant 10^{38}\, \rm erg/s$ 
    will improve with time, their confirmation, however, could be significantly reduced due to the contamination from XRBs.

\end{itemize}

The BH and AGN occupation fractions have implications for the evolution of low-mass galaxies in the simulations. For example, the small BH OF in low-mass galaxies at $z\leqslant 2$ in Horizon-AGN prevents these galaxies from any energy release via AGN feedback. The role of AGN feedback in low-mass galaxies is still unclear today \citep[e.g.,][]{2018MNRAS.473.5698D,2019MNRAS.484.2047K}, but evidence for feedback has been observed \citep[e.g.,][]{2019MNRAS.488..685M,2019ApJ...884...54M}.
Theoretically, the effect could be small as the BHs have low masses and could have reduced accretion rates due to SN feedback \citep{2015MNRAS.452.1502D,2017MNRAS.468.3935H,2017MNRAS.472L.109A}, but AGN feedback could also provide a second channel to clear out the low-mass galaxies from their gas content and enhance the effect of SN feedback.
On the other hand, if too many (and/or too bright) AGN are produced in the simulations compared to observations, they can potentially release too much energy in the host galaxy history, which can e.g., alter the properties of the gas, and galaxy gas and stellar contents. In the hierarchical build-up of galaxies, this can  potentially affect the initial stages of 
many galaxies, and thus needs to be considered and further investigated in the large-scale cosmological simulations.

\section*{Acknowledgments}
We are grateful to the anonymous referee for their insightful comments and suggestions that improved this work. Authors would also like to thank G. Mamon, S. Kaviraj, and S. Genel for fruitful discussions.
HH thanks the Institut d'Astrophysique de Paris for hosting her. 
M.M. acknowledges support from the Ramon y Cajal fellowship (RYC2019-027670-I).
DAA was supported in part by NSF grants AST-2009687 and AST-2108944, and by the Flatiron Institute, which is supported by the Simons Foundation.

\section*{Data Availability Statement}
The data from the Illustris and the TNG100 simulations can be found on their respective websites: https://www.illustris-project.org, https://www.tng-project.org. The data from the EAGLE simulation can be obtained upon request to the EAGLE team at their website: http://icc.dur.ac.uk/Eagle/. 
The data from the SIMBA simulation can be found on the website: http://simba.roe.ac.uk/.
The Horizon-AGN simulation is not public, but some catalogs are available at: https://www.horizon-simulation.org/data.html.

\bibliographystyle{mn2e}
\bibliography{biblio_complete_faintQSOs,biblio_complete,biblio_complete-2,reference}

\appendix
\section{Current observational constraints on the AGN occupation fraction}
\label{sec:observations_list}

We provide in this section an exhaustive list of existing constraints on the AGN OF in the low-redshift Universe. These observational constraints are used in Section~\ref{sec:agn_occ_fraction} and reported in Fig.~\ref{fig:AGN_obs_vs_sim}.

The early work of \citet{kauffmann2003host} derive the AGN OF from the SDSS survey, finding a fraction below $10\%$ for galaxies with $M_{\star}\leqslant 10^{10}\, \rm M_{\odot}$, which peaks at about $25\%$ in galaxies with $M_{\star}\sim 10^{11}\, \rm M_{\odot}$.

The constraints of \citet{2008ApJ...688..794S} 
use a luminosity threshold of $L_{\rm x,\, AGN, \, 2-8\, keV}\geqslant 10^{42}\, \rm erg/s$, and galaxies with $M_{\star}=5\times 10^{9}- 2\times 10^{10}\, \rm M_{\odot}$. The AGN of this sample have 2-8 keV luminosities mostly in the range $L_{\rm x,\, AGN, \, 2-8\, keV}=10^{42}-10^{43.5}\, \rm erg/s$. This can be converted to the $2-10\, \rm keV$ band using $L_{2-8\, \rm keV} /L_{2-10\, \rm keV} = 0.86$, assuming a power law photon index of 1.0 \citep{2008ApJ...688..794S}. As such, we get: $L_{\rm x,\, AGN, \, 2-10\, keV}=10^{41.9}-10^{43.4}\, \rm erg/s$.

\citet{Aird2012} 
identify more than 200 AGN with $ L_{\rm x,\, AGN, \, 2-10 keV}\geqslant 10^{42}\, \rm erg/s$ in more than 25000 galaxies at $z<1$. For galaxies with $ M_{\star}\sim 10^{9.75}\, \rm M_{\odot}$ and $0.2<z<0.6$, the occupation fractions are within the ranges OF$= 0.3-1.6, 0.17-0.5, 0.01-0.17\, \%$ for X-ray (2-10 keV) AGN luminosities of $L_{\rm x}=10^{42}-10^{42.5}, 10^{42.5}-10^{43}, 10^{43}-10^{43.5}\, \rm erg/s$, respectively. 
For $0.6<z<1$ and galaxies with $ M_{\star}\sim 10^{10.3}\, \rm M_{\odot}$, the authors find AGN occupation fractions of $\rm OF= 0.6-1.6, 0.3-1.0, 0.2-0.3\, \%$ for $L_{\rm x}=10^{42}-10^{42.5}, 10^{42.5}-10^{43}, 10^{43}-10^{43.5}\, \rm erg/s$, respectively.  
\citet{Aird2012} find that at fixed AGN luminosity the AGN fraction decreases toward less massive galaxies in the range $M_{\star}=10^{10}-10^{12}\, \rm M_{\odot}$. If this trend extends down to even lower mass galaxies, the AGN OF in galaxies of $M_{\star}< 10^{9.75}\, \rm M_{\odot}$ could be lower than the AGN fractions listed above. 

\citet{schramm2013black} report X-ray detections in more than 5000 galaxies with $M_{\star}\leqslant 3\times 10^{9}\, \rm M_{\odot}$ at $z<1$, among which 3 detections out of 2100 galaxies at $z<0.3$. The latter provides a fraction of $ \rm OF=0.14\%$ at $z<0.3$. These fractions are for a galaxy X-ray luminosity of $L_{\rm x, \, gal, \, 0.5-8\, keV}\leqslant 10^{44}\, \rm erg/s$.
The three detections are in galaxies with $L_{\rm x, \, gal, \, 0.5-8\, keV}=10^{40.3}, 10^{40.3}, 10^{41.1}\, \rm erg/s$. The SFR and $M_{\star}$ of these three galaxies are given in \cite{schramm2013black}, which allows us to estimate the AGN luminosities by subtracting the contribution of the XRBs and the ISM.

The constraints of \citet{reines2013dwarf} include 
$\sim 150$ 
galaxies that have been identified with BH accretion signatures out of $25974$ galaxies, i.e. 136 galaxies with photoionization signatures of BH accretion (and for some of them broad $H\alpha$ emission) and 16 galaxies with narrow-line ratios which are consistent with AGN or star formation. Based on the follow-ups carried by \citet{baldassare2016multi} on the $16$ candidates, $11$ galaxies had a broad $\rm H_{\alpha}$ line that vanished and were thus discarded as AGN, $2$ galaxies were proven to be AGN, and $3$ galaxies were still ambiguous and still qualified as AGN candidates. From the $140$ remaining galaxies, we derive an AGN fraction of $\rm OF=0.53\%$ 
for galaxies with $M_{\star}\leqslant 10^{9.5}\, \rm M_{\odot}$ in the local Universe. From the X-ray follow up of \citet{2017ApJ...836...20B}, the hard X-ray luminosity of these AGN are within the range $L_{\rm x, \, AGN, \, 2-10\, keV}=10^{39.76}- 10^{41.84}\, \rm erg/s$.

\citet{2014ApJ...784..113S} investigate the presence of AGN in 13862 low-redshift bulgeless galaxies from SDSS with extremely red mid-IR colors, possible signature of an AGN. From 25 to 300 AGN candidates are extracted with the color selection; they do not have AGN signature in the optical. The AGN fraction is about $\leqslant 40\%$ in bulgeless galaxies with $M_{\star}=10^{8}-10^{10}\, \rm M_{\odot}$.

\citet{moran2014black} finds an OF of $2.7\%$ for galaxies with $M_{\star}=4\times 10^{9}-10^{10}\, \rm M_{\odot}$. The study uses a distance-limited sample of galaxies from SDSS, in which 28 AGN are identified based on their optical emission lines.

\citet{lemons2015x} identify $19$ galaxies with X-ray detections out of about $491$ galaxies with $M_{\star}\leqslant 3\times 10^{9}\, \rm M_{\odot}$ \citep[here we used the estimate of the total number of galaxies from][]{pardo2016x}.
In the sample, 10 galaxies have signatures of a central AGN, and 8 galaxies have a higher X-ray emission than the predicted galaxy-wide emission from XRBs. From the latter, we estimate an AGN fraction of $\rm OF=1.63\%$. X-ray sources have $2-10\, \rm keV$ luminosities of $L_{\rm x,\, source,\, 2-10\, keV}\leqslant 10^{40.1}\, \rm erg/s$.

\citet{2015ApJ...799...98M} find that $1-2\%$ of galaxies with $M_{\star}\leqslant 10^{9}\, \rm M_{\odot}$ have nuclear X-ray emission. We consider the result as an upper limit, since XRBs could contaminate the emission. We also report these results in all the panels of Fig.~\ref{fig:AGN_obs_vs_sim}, since no X-ray luminosities are given in this study.

\citet{2015MNRAS.454.3722S} identify 336 AGN candidates in a sample of $\sim 48 000$ SDSS dwarf galaxies with $M_{\star}\leqslant 10^{9.5}\, \rm M_{\odot}$ at $z<0.1$. AGN are selected based on the BPT diagram, but also if they exhibit optical narrow-line signatures of an accreting BH (He II selected AGN), or based on galaxy mid-IR colors consistent with the presence of an AGN.
The AGN OF is within the range $\sim 0.02-0.5\%$ for galaxies with $M_{\star}\sim 10^{8.5}\, \rm M_{\odot}$, and of $0.1-0.3\%$ for galaxies with $M_{\star}\sim 10^{9.5}\, \rm M_{\odot}$ and for these three selection methods.

\citet{pardo2016x} derive an occupation fraction of $\rm OF=0.6-3\%$ for galaxies with $M_{\star}=10^{9}-3\times 10^{9}\, \rm M_{\odot}$, and $L_{\rm x,\, source, 0.5-7\, keV}=10^{39.9}-10^{42.1}\, \rm erg/s$. From the hardness ratios, the authors found a low probability of contamination from X-ray emitted by HMXBs and/or from the hot gas associated with SN remnants. We report these values in Fig.~\ref{fig:AGN_obs_vs_sim}, although the band used in \citet{pardo2016x} is not hard 2-10 keV X-ray as used in this paper.

Furthermore, we include the constraints of \cite{2016ApJ...817...20M} which investigate the presence of AGN in dwarf starburst and spiral galaxies up to $z=1.5$. They use stacking X-ray counts from more than 7000 of galaxies in the range $0<z<0.3$, and about 10000 galaxies and 3000 galaxies in the range $0.7<z<1.0$, and $1<z<1.5$, respectively. 
The fraction of X-ray excess after removing the possible contributions from XRBs and hot ISM is $\sim 70\%$ in the redshift bins below $z<0.7$, and is $\sim 50\%$ in the bins of higher redshifts. From stacking analysis, we can not assess whether the X-ray excess is due to a few galaxies with powerful AGN, or to many galaxies with more moderate AGN. However, the results of \cite{2016ApJ...817...20M} suggest that AGN could be common in low-mass galaxies up to $z=1.5$.

\citet{2017A&A...602A..28M} investigate the presence of AGN in dwarf galaxies (62) with stellar masses from $10^{6}$ to $10^{9}\, \rm M_{\odot}$. Their sample consists of 303 candidate AGN, selected based on IR colors.
Using  the MPA/JHU catalogue, \citet{2017A&A...602A..28M} derives an AGN fraction that does not exceed $5\%$ across the stellar mass range $10^{6} \leqslant M_{\star}/\, \rm M_{\odot} \leqslant 10^{11}$. 
The fraction slightly increases as a function of decreasing stellar mass for $M_{\star} \leqslant 10^{9}\, \rm M_{\odot}$.

\cite{2018MNRAS.478.2576M} constrains the AGN fraction in galaxies with $M_{\star}=10^{9}-3\times 10^{9}\, \rm M_{\odot}$.
The fraction of AGN in dwarf galaxies is $\rm OF\sim 0.43\%$ for $L_{\rm x,\, AGN,\, 0.5-10\,keV}=10^{41.5}-10^{42.4}\, \rm erg/s$ and at $z<0.3$, and $\rm OF\sim 0.09\%$ for $L_{\rm x,\, AGN,\, 0.5-10\,keV}\geqslant 10^{42.4}\, \rm erg/s$ and at $z<0.3$.
At higher redshift of $z\sim 0.6$, the AGN OF is lower with $\rm OF\sim 0.04\%$ for $L_{\rm x,\, AGN,\, 0.5-10\,keV}\geqslant 10^{42.4}\, \rm erg/s$.

\citet{2018MNRAS.474.1225A} find an AGN fraction of $\rm OF\sim 0.1-0.9\%$ in galaxies with $M_{\star}=10^{8.5}-10^{9.5}\, \rm M_{\odot}$. This fraction is defined as the number of AGN with an Eddington ratio $f_{\rm Edd}>0.01$. The exact definition in \citet{2018MNRAS.474.1225A} is based on AGN with $k_{\rm bol}\times L_{\rm x,\,AGN, \, 2-10\, keV}/(1.38\times 10^{38}\,{\rm erg/s}\times 0.002\times M_{\star}/{\rm M_{\odot}})>0.01$, with this quantity being similar to $f_{\rm Edd}$ for their choice of parameters. This limit can be converted into a limit in 2-10 keV luminosity of $L_{\rm x,\,AGN,\, 2-10\, keV}\geqslant 10^{41}-10^{41.5}\, \rm erg/s$ for the range $M_{\star}=10^{9}-10^{9.5}\, \rm M_{\odot}$.

\citet{2018ApJ...863....1C} provides a  sample of 305 AGN candidates with Type 1 signatures in the range $M_{\rm BH} = 3\times 10^{4}- 2\times 10^{5} \, \rm M_{\odot}$ in galaxies with $M_{\rm bulge}=10^{7}-10^{12}\, \rm M_{\odot}$, out of which 10 sources were confirmed as AGN. 
The X-ray luminosity of the AGN varies in $L_{\rm x, \, AGN}=10^{40}-3\times 10^{42}\, \rm erg/s$, for different X-ray bands.

\cite{dickey2019agn} present constraints on the AGN fraction in a small sample of galaxies, including 20 quiescent galaxies and 7 star-forming galaxies. Interestingly, many of the quiescent galaxies show signatures of AGN activity (from optical emission line ratios), and leads to an occupation fraction of $\rm OF=80\%$ in those galaxies. A fraction of $\rm OF=28.5\%$ is found in the star-forming galaxies. Given the small sample studied in \cite{dickey2019agn}, we do not add these constraints to Fig.~\ref{fig:AGN_obs_vs_sim}.

\citet{2019MNRAS.489L..12K} recently derived the fraction of AGN in a wide range of galaxy stellar mass ($M_{\star}=10^{7}-10^{11.5}\, \rm M_{\odot}$), using diagnostics based on IR colors. The AGN fraction varies in the range $\rm OF=20-40\%$ for their fiducial model and $M_{\star}=10^{9}-10^{10.5}\, \rm M_{\odot}$, and $\rm OF=10-25\%$ in the same galaxy mass range when adopting other color criteria that have been used in the literature. An estimate of the hard X-ray luminosity of these AGN is $L_{\rm AGN}\leqslant 10^{42}\, \rm erg/s$ (S. Kaviraj, private communication).

\citet{2020MNRAS.492.2528L} recently used the same sample of \citet{2019MNRAS.489L..12K} to find that 7 to 20 candidates could be AGN, over 5000 galaxies with $M_{\star}=10^{8}-10^{9}\, \rm M_{\odot}$. This corresponds to an OF of $0.14-0.4\, \%$ for AGN with estimated bolometric luminosities of $L_{\rm bol, \, AGN}\leqslant 3\times 10^{44}\, \rm erg/s$. This is lower than the $\geqslant 10\,\%$ fractions found in \citet{2019MNRAS.489L..12K}. \citet{2020MNRAS.492.2528L} convoluted both optical and mid-IR data and took into account the contamination from close sources, the relatively low resolution of the WISE mid-IR instrument that was used (with respect to for example optical surveys), and the contamination from the galaxies. The difference among the two studies demonstrates the difficulty of using mid-IR selection to identify AGN in these galaxies.

\citet{2020MNRAS.492.2268B} investigate the fraction of AGN (detected in X-ray, 2-12 keV) in different stellar mass ranges of galaxies in the local Universe ($z\leqslant 0.25$). For $M_{\star}=10^{7}-10^{9}\rm \, M_{\odot}$, the AGN fraction is $\rm OF\sim 0.19\%$ 
for $L_{\rm x,\, AGN, \, 2-12\, keV}=10^{41.5}-10^{42.4}\, \rm erg/s$, and $\rm OF\sim 0.1\%$ 
for $L_{\rm x,\, AGN, \, 2-12\, keV}\geqslant 10^{42.4}\, \rm erg/s$. For more massive galaxies with $M_{\star}=10^{9}-10^{9.5}\,\rm M_{\odot}$, \citet{2020MNRAS.492.2268B} find a fraction of $\rm OF\sim 0.17\%$ 
and $\rm OF\sim 0.06\%$, 
for AGN with $L_{\rm x,\,  AGN, \, 2-12\, keV}=10^{41.5}-10^{42.4}\, \rm erg/s$ and $L_{\rm x,\, AGN, \, 2-12\, keV}\geqslant 10^{42.4}\, \rm erg/s$ respectively.

\citet{2020ApJ...898L..30M} identify 37 galaxies over a sample of 1609 dwarf galaxies with $M_{\star}\leqslant 10^{9.5}\, \rm M_{\odot}$ with signatures consistent with AGN ionisation. This results in a fraction of $\rm OF=2.3\%$ for AGN in the range $L_{\rm bol, \, AGN}=10^{38.9}-10^{41.4}\, \rm erg/s$. We convert these values into hard X-ray luminosities
using the bolometric correction of \citet{2007ApJ...654..731H}.

\citet{2021arXiv211110388L} recently studied the AGN fraction in galaxies with $M_{\star}\leqslant 3\times 10^{9}\, \rm M_{\odot}$ using X-ray observations from the eROSITA eFEDS survey, for $z\leqslant 0.15$. The authors derived an upper limit to the AGN OF of the sample of $\leqslant 1.8\,\%$. The sources have X-ray luminosities in the range $L_{\rm x, \, gal, \, 2-10\, keV}=10^{38.5}-10^{40}\, \rm erg/s$.

Finally, \citet{2022MNRAS.510.4556B} recently reported the AGN fraction in stellar mass bins for luminous AGN with $L_{\rm x,\,AGN,\, 2-10\, keV }=10^{42}-10^{44}\, \rm erg/s$ derived from the MPA-JHU catalogue (based on SDSS DR8) and 3XMM.
The AGN fraction increases towards more massive galaxies, from $3-9\times 10^{-4}$ in galaxies with $\leqslant 10^{9.5}\, \rm M_{\odot}$ to $2-6 \times 10^{-3}$ in $\sim 10^{10.5}\,\rm M_{\odot}$ galaxies. At fixed stellar mass, the fraction also slightly increases towards higher redshifts in the range $z=0-0.3$.

To summarize, the measured AGN OF depends on the different studies, selection techniques, and sample definitions. Most of the constraints listed in this section span over an order of magnitude or more, but in general favor an OF of $\leqslant 3\times 10^{-2}$ in galaxies of $M_{\star}=10^{8.5}-10^{10}\, \rm M_{\odot}$. 
We emphasize here that only a few of these analyses include a correction for galaxy completeness 
\citep{Aird2012,2015ApJ...799...98M,pardo2016x,2018MNRAS.474.1225A,2018MNRAS.478.2576M,2020MNRAS.492.2268B}, that can mostly be done with deep X-ray surveys.
In theory, deriving an OF requires to weigh each AGN detection by the completeness of the corresponding bin (defined by a redshift, galaxy stellar mass, and X-ray luminosity). The weighting function thus depends on the total number of galaxies which should be in the bin, the X-ray sensitivity of the survey (i.e., what is the probability of detecting an AGN of a given luminosity in the survey, the probability varies spatially). We refer the reader to the Section of \citet{Aird2012} for an in-depth description of the methodology.

\section{Simulated AGN in the low-mass galaxy regime}
\label{sec:AGN_lum_distri}
\begin{figure*}
    \hspace*{-0.5cm}
    \includegraphics[scale=0.255]{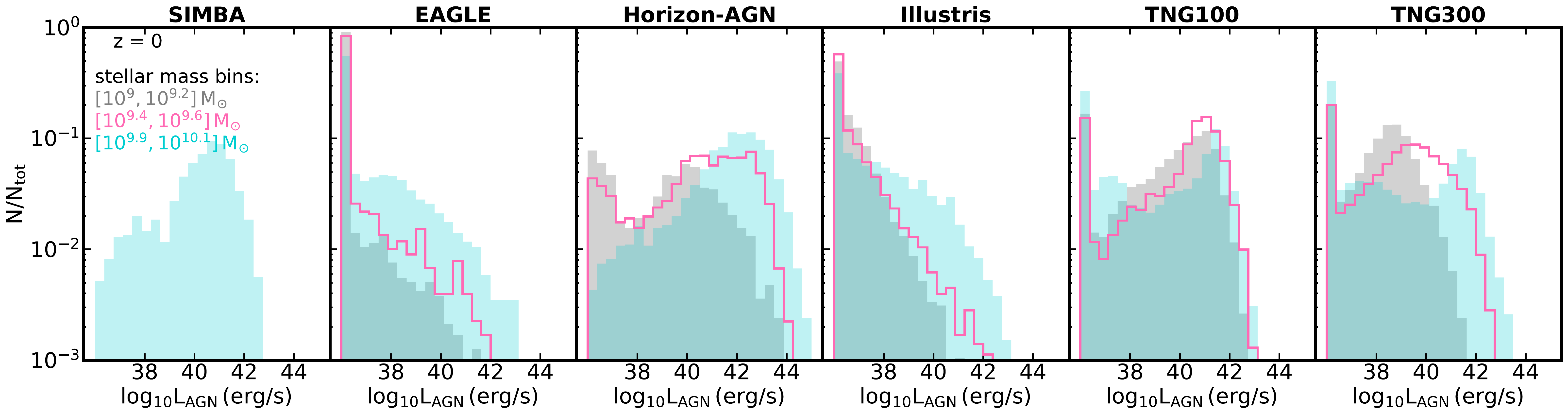}
    \caption{The normalised distribution of X-ray AGN luminosity at $z=0$ galaxies with $M_{\star}\sim 10^{9},\, \sim 10^{9.5}, \, \sim 10^{10}\, \rm M_{\odot}$ for all the six simulations. For SIMBA, we do not display the BH population for $M_{\star} <10^{9.5} \rm M_{\odot}$ since the simulation starts seeding BHs at the aforementioned value. 
    }
    \label{fig:Lxray-hist}
\end{figure*}

Besides having different AGN occupation fractions at $z=0$, the simulations have significant differences in their AGN populations. We show the histograms of the AGN luminosities in Fig.~\ref{fig:Lxray-hist} for AGN in galaxies with total stellar mass of $\sim 10^{9}$, $\sim 10^{9.5}$, and $10^{10}\, \rm M_{\odot}$.
The AGN luminosity distribution in galaxies with $M_{\star}\sim 10^{10}\, \rm M_{\odot}$ peaks at $\log_{10}\, L_{\rm x}/(\rm erg/s )\sim 41$ in SIMBA (median $\log_{10}\, L_{\rm x}/(\rm erg/s)\sim 39$), and at $\log_{10}\, L_{\rm x}/(\rm erg/s) \sim 42$ for Horizon-AGN (median $\log_{10}\, L_{\rm x}/(\rm erg/s)\sim 42$), TNG100 (median $\log_{10}\, L_{\rm x}/(\rm erg/s)\sim 40$), and TNG300. However, we find that the EAGLE and Illustris simulations have fainter AGN populations in these intermediate-mass galaxies.
In the dwarf regime, i.e. for galaxies with $M_{\star}\sim 10^{9}\, \rm M_{\odot}$, AGN are on average fainter for most of the simulations. In Horizon-AGN, the AGN luminosity distribution has a median of $\log_{10}\, L_{\rm x}/(\rm erg/s)\sim 37$, $\log_{10}\, L_{\rm x}/(\rm erg/s)\sim 31$ in EAGLE, and  $\log_{10}\, L_{\rm x}/(\rm erg/s)\sim 37$ in Illustris. However, we find that the AGN in the dwarf galaxies produced by TNG100 still have similar luminosities (median $\log_{10}\, L_{\rm x}/(\rm erg/s)\sim 40$) than in more massive galaxies with $M_{\star}\sim 10^{10}\, \rm M_{\odot}$.

Assessing the agreement between the AGN luminosity functions produced by the simulations and those constrained by observations is crucial. A comparison is presented in \citet{2022MNRAS.509.3015H} and shows that at $z=0$ the faint end of the hard X-ray luminosity function ($L_{\rm x}\leqslant 10^{43}\, \rm erg/s$) of Horizon-AGN, Illustris, and SIMBA are within the constraints of \citet{2015ApJ...802...89B} and in agreement with those of \citet{Hop_bol_2007,2015MNRAS.451.1892A}. The function produced by TNG100 is above the constraints, and the one from EAGLE slightly below. A high number of faint AGN at $z=0$ are located in low-mass galaxies in TNG100 and could explain the higher luminosity function in case of too massive BHs, and/or too efficiently accreting BHs, and/or not strong enough SN feedback compared to observations. In the opposite, the lower luminosity function produced in EAGLE could be explained by lower-mass BHs than in observations (a priori not the case, see Fig.~\ref{fig:mgal-mbh-scatter}) and/or too weak accretion onto the BHs and/or too efficient SN feedback in low-mass galaxies \citet{2022MNRAS.509.3015H}.

\section{Contributors to galaxy X-ray luminosity}
\label{sec:XRAY-contribution}

\begin{figure}
    \centering
    \includegraphics[scale=0.5]{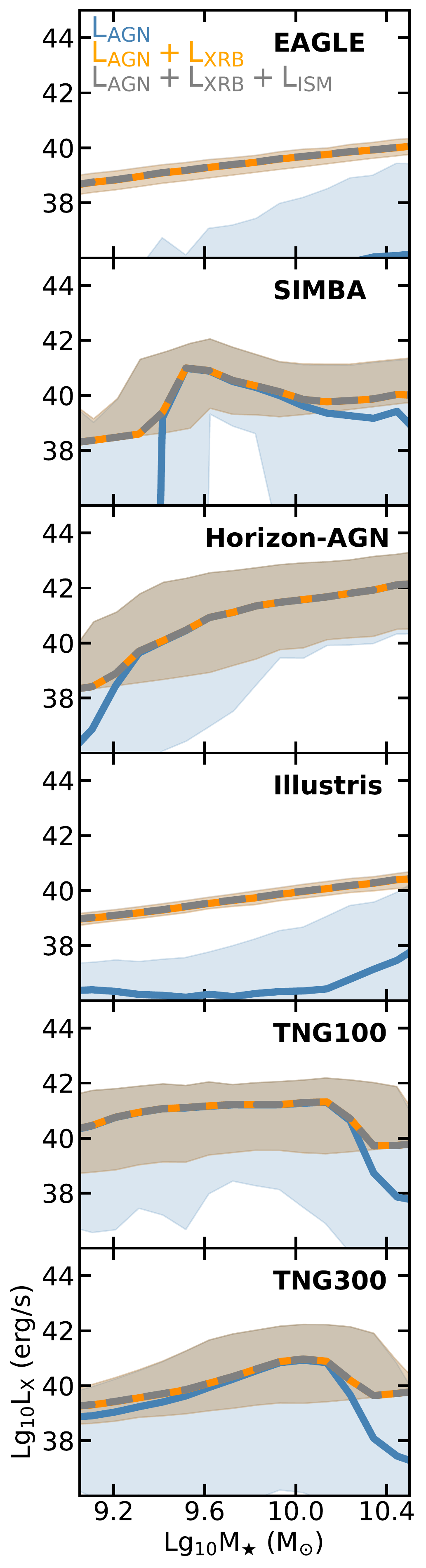}
     \caption{Median of the hard X-ray luminosity as function of stellar mass in the low-mass regime. We display three contributors to the galaxy X-ray luminosity: $L_{\rm AGN}$ (blue), $L_{\rm AGN} + L_{\rm XRB}$ (orange) and $L_{\rm AGN} + L_{\rm XRB} + L_{\rm ISM}$ (grey). The 15-85th percentiles are illustrated by the shaded areas.}
    \label{fig:different_xray_vs_mgal}
\end{figure}

In this paper, we use the hard X-ray luminosity of the simulated AGN to assess whether they could be detected, and whether the XRB population could outshine them.
However, observations often rely on a galaxy-wide X-ray emission, which is the sum of the AGN emission, the XRB population, and the ISM host gas. Using the X-ray stacking technique, \citet{2012ApJ...753L..30M} and \citet{2018MNRAS.474.1225A} compute the average AGN luminosity in star-forming galaxies at $z\sim 1$ and $z\sim 2$ and find that the $<L_{AGN}>-M_{\star}$ relation behave linearly (also referred to as ``AGN main sequence''). This is not always true for low-mass galaxies, where \citet{2018MNRAS.474.1225A} find that AGN activity is significantly suppressed. 
In the following, we show that, in the absence of obscuration, the emission from the AGN in the low-mass end $M_{\star} \leqslant 10^{10.5}\, \rm M_{\odot}$ is on average larger than the XRBs and ISM for most of the simulations. The average contributions of the AGN, XRB population, and ISM to the galaxy-wide X-ray luminosity are shown in Fig.~\ref{fig:different_xray_vs_mgal}.

The contribution of the AGN differs from a simulation to another. AGN in EAGLE are on average the faintest followed by Illustris, both which have a median hard X-ray luminosity $L_{\rm AGN} \leqslant 10^{38}\, \rm erg/s$. 
The relative contribution of the XRB population is higher for these simulations at all stellar masses, and is higher in the other simulations in regimes where the AGN luminosity is lower: typically dwarf galaxies with $\sim 10^{9}\, \rm M_{\odot}$ (e.g., Horizon, TNG300), and more massive galaxies with $M_{\star}\geqslant 10^{10.2}\, \rm M_{\odot}$ in TNG when the efficient kinetic AGN feedback mode kicks out.
In these regimes, the XRB population is able to outshine the AGN.
We find that the contribution from ISM to the galaxy-wide X-ray luminosity is very small for all the simulations.

\section{AGN and galaxy-wide XRB luminosities}

In Table~\ref{tab:table-percentage-galtype-AGN}, we provide three different percentages 
for three different galaxy types (starburst, main sequence and quenched), and for three stellar mass bins: $\sim 10^{9}, 10^{9.5}$ and $10^{10}\, \rm M_{\odot}$:
\begin{itemize}
    \item The percentage of galaxy types in each simulation, referred as ``starburst'', ``Main Seq'' and ``Quenched'' in Table~\ref{tab:table-percentage-galtype-AGN}. This is defined as the number of galaxies of a given galaxy type in a given stellar mass bin, divided by the total number of galaxies in that bin.
    \item The percentage of galaxies, within a given galaxy type, hosting an AGN with $L_{\rm AGN} \geqslant 10^{38}\, \rm erg/s$.

    \item The percentage of galaxies, within a given galaxy type, that host an AGN with $L_{\rm AGN} \geqslant 10^{38}\, \rm erg/s$ and satisfy $L_{\rm AGN} > L_{\rm XRB}$ (i.e., an AGN that outshines the galaxy-wide XRB population).

\end{itemize}

Our definitions of the galaxy types are simulation dependent.
We also provide the percentages for AGN with $L_{\rm AGN} \geqslant 10^{40}\, \rm erg/s$.

\begin{table*}
\centering
\caption{Percentages of galaxy types (starburst, main sequence and quenched) at $z=0$ in three different stellar bins: $\log_{10}M_{\star}/{\rm M_{\odot}} \sim 9,\, 9.5,\, 10$. 
For each galaxy type, we derived the percentage of galaxies that host an AGN with a hard X-ray luminosity $L_{\rm AGN} \geqslant 10^{38} \, \rm erg/s$ and $L_{\rm AGN} \geqslant 10^{40} \, \rm erg/s$. We also add the percentages of galaxies with an AGN outshining the galaxy-wide XRB population.}

\begin{tabular}{|l|l|l|l|l|l|l|l|}
\hline
$\log_{10}M_{\star}/\rm M_{\odot}$ & $\%$ of & EAGLE & SIMBA & Horizon-AGN & Illustris & TNG100 & TNG300 \\
\hline

9-9.2 & Starburst &6 &-& 4&2&4&4\\
& AGN host $(L_{\rm AGN} > 10^{38})$  & 6 &- &2&30&
89&80 \\
& $L_{\rm AGN} > 10^{38}$ and $L_{\rm AGN} > L_{\rm XRB}$   & 3 &- &1&10&80&42 \\
& AGN host $(L_{\rm AGN} > 10^{40})$  & 2 &- &1&7&78&27 \\
& $L_{\rm AGN} > 10^{40}$ and ${\rm AGN} > L_{\rm XRB}$   & 2 &- &1&7&78&27 \\\cline{2-8}

&  Main Seq. & 61&- &75&86&74&70 \\
& AGN host $(L_{\rm AGN} > 10^{38})$  & 4 &- &3&6&82&64 \\
& $L_{\rm AGN} > 10^{38}$ and $L_{\rm AGN} > L_{\rm XRB}$   & 2 &- &2&0&70&22 \\
&  AGN host$(L_{\rm AGN} > 10^{40})$  &1 &-&2&1&50&5\\
& $L_{\rm AGN} > 10^{38}$ and $L_{\rm AGN} > L_{\rm XRB}$   & 1 &- &2&1&50&5\\\cline{2-8}
&Quenched &33 &- &21&12&22&26 \\
& AGN host $(L_{\rm AGN} > 10^{38})$  & 0 &- &3&0&1&0 \\
& $L_{\rm AGN} > 10^{38}$ and $L_{\rm AGN} > L_{\rm XRB}$   & 0 &- &2&0&0&0 \\
&  AGN host $(L_{\rm AGN} > 10^{40})$ & 0 &- &1&0&0&0 \\
& $L_{\rm AGN} > 10^{40}$ and $L_{\rm AGN} > L_{\rm XRB}$   & 0 &- &1&0&0&0 \\\cline{2-8}

 \hline
9.4-9.6& Starburst & 5&- &4&1&2&5  \\
& AGN host $(L_{\rm AGN} > 10^{38})$  &12  &- &7&38&93&87 \\
& $L_{\rm AGN} > 10^{38}$ and $L_{\rm AGN} > L_{\rm XRB}$   & 6 &- &7&20&80&60 \\
& AGN host $(L_{\rm AGN} > 10^{40})$ &5 &- &7 &21&85&57 \\
& $L_{\rm AGN} > 10^{40}$ and $L_{\rm AGN} > L_{\rm XRB}$   & 5 &- &7&21&85&56 \\\cline{2-8}

& Main Seq. &73 &- &82&91&78& 71 \\
& AGN host$(L_{\rm AGN} > 10^{38})$  & 9 &- &18&10&88&74 \\
& $L_{\rm AGN} > 10^{38}$ and $L_{\rm AGN} > L_{\rm XRB}$   & 4 &- &16&2&70&46 \\
& AGN host $(L_{\rm AGN} > 10^{40})$ &3 &- &15&2&69&29 \\
& $L_{\rm AGN} > 10^{40}$ and $L_{\rm AGN} > L_{\rm XRB}$   & 3 &- &15&2&69&29\\\cline{2-8}

& Quenched &23 &- &13&8&20&24 \\
& AGN host $(L_{\rm AGN} > 10^{38})$  &0  &- &12&0&3&1 \\
& $L_{\rm AGN} > 10^{38}$ and $L_{\rm AGN} > L_{\rm XRB}$   & 0 &- &10&0&0&0 \\
&  AGN host$(L_{\rm AGN} > 10^{40})$ &0 &- &6&0&0&0 \\
& $L_{\rm AGN} > 10^{40}$ and $L_{\rm AGN} > L_{\rm XRB}$   & 0 &- &6&0&0&0 \\\cline{2-8}

\hline
9.9-10.1& Starburst &2& 4&4&1&3&5 \\
& AGN host $(L_{\rm AGN} > 10^{38})$  & 27 &99&28&75&100&92 \\
& $L_{\rm AGN} > 10^{38}$ and $L_{\rm AGN} > L_{\rm XRB}$   & 18 &98&7&50&90&76 \\
& AGN host$(L_{\rm AGN} > 10^{40})$  &18&98&27&58&93&81\\
& $L_{\rm AGN} > 10^{40}$ and $L_{\rm AGN} \geq L_{\rm XRB}$   &5  &98 &27&42&90&76 \\\cline{2-8}

& Main Seq. &76&59&87&92&79&73 \\
& AGN host $(L_{\rm AGN} > 10^{38})$ &19  &95&50&22&93&85 \\
& $L_{\rm AGN} > 10^{38}$ and $L_{\rm AGN} > L_{\rm XRB}$   & 7 &64&17&6&71&69 \\
& AGN host $(L_{\rm AGN} > 10^{40})$ &6&64&45&6&74&65 \\
& $L_{\rm AGN} > 10^{40}$ and $L_{\rm AGN} \geq L_{\rm XRB}$   & 2 &64 &45&6&74&64 \\\cline{2-8}

& Quenched &22&38&9&8&18&22 \\
& AGN host $(L_{\rm AGN} > 10^{38})$  & 3 &30&27&0&4&7 \\
& $L_{\rm AGN} > 10^{38}$ and $L_{\rm AGN} > L_{\rm XRB}$   & 2 &14&10&0&0&1 \\
&  AGN host $(L_{\rm AGN} > 10^{40})$  &2&6&17&0&2&1 \\
& $L_{\rm AGN} > 10^{40}$ and $L_{\rm AGN} \geq L_{\rm XRB}$   & 0 &5 &17&0&2&0\\\cline{2-8}

 \hline
\end{tabular}
\label{tab:table-percentage-galtype-AGN}
\end{table*}

\section{Observational samples presented in Fig.~7}

\label{sec:obs_SFR_Mgal}

We describe in this section the observational samples presented in Fig.~\ref{fig:sfr-mgal-scatter}.
We use the nearby host galaxies ($z<0.5$) of \citet[][see Appendix~\ref{sec:observations_list} for details]{2018MNRAS.478.2576M}
where the AGN luminosity ranges from $L_{\rm x,\, AGN,\,0.5–10 keV} \sim 10^{39} - 10^{43} \, \rm erg/s$ and the SFR of their hosts from $\rm \log_{10} SFR/(M_{\odot}/yr) = [-2.63,1.11]$. We also show 37 AGN candidates of \citet{2020ApJ...898L..30M}, 
identified in galaxies with SFR $\rm \log_{10} SFR/(M_{\odot}/yr) = -2.5, 1.5$. For a fair comparison, we have converted their AGN luminosities from bolometric to hard (2-10 keV) X-ray \citep[using the correction of][]{Hop_bol_2007}.

We also include the sample of \citet{2020MNRAS.492.2268B} 
presenting 61 dwarf galaxies as potential AGN hosts 
with SFR in the range $\rm \log_{10} SFR/(M_{\odot}/yr) = -3.8, 1.2$. 
The X-ray luminosity of this work are in the $2-12$ keV band and for the galaxy as a whole.
To derive the luminosities of the AGN (as shown in Fig.~\ref{fig:sfr-mgal-scatter}), we subtract the predicted contribution of the XRB population \citep[using the relation of][]{2016ApJ...825....7L} and hot gas from the ISM \citep[using the relation of][]{2012MNRAS.419.2095M} from the X-ray galaxy luminosity given in their sample.

For our analysis, we search for X-ray counterparts to the AGN of \citet{2021MNRAS.tmpL..99M} using the Chandra Source Catalog \citep{2010ApJS..189...37E} version 2 (CSC 2.0) Master Source Catalog. We use a search radius of $5 \, \rm arcsec$, which returns 50 X-ray detections. For these, we take from CSC 2.0 the aperture-corrected net energy flux, inferred from the source region aperture, in the ACIS $2-7.0\, \rm keV$ energy band. This flux corresponds to the best estimate derived from the longest block of a multi-band, flux-ordered Bayesian Block analysis of the contributing observations. The K-corrected X-ray luminosity in the $2-10\, \rm keV$ band is derived from this aperture-corrected flux assuming a typical AGN photon index of $1.8$.

\label{lastpage}

\end{document}